\documentclass[twocolumn]{aastex61}
\usepackage{comment}
\usepackage{amsmath,amssymb}
\usepackage{graphicx}
\usepackage{textcomp}
\usepackage{booktabs}
\usepackage{bm}
\usepackage{float} 

\newcommand\aastex{AAS\TeX}

\newcommand\ltsima{$\; \buildrel <\over\sim \;$}
\newcommand\simlt{\lower.5ex\hbox{\ltsima}}
\newcommand\gtsima{$\; \buildrel >\over\sim \;$}
\newcommand\simgt{\lower.5ex\hbox{\gtsima}}

\shorttitle{\aastex\ PRIME: First Results}
\shortauthors{Sumi et al.}
\defcitealias{Koshimoto2023}{K23}
\defcitealias{Koshimoto2021}{Koshimoto+21a}
\defcitealias{Gould2022}{Gould+22}

\begin{document}

\title{The Prime Focus Infrared Microlensing Experiment (PRIME): First Results}

\author[0000-0002-4035-5012]{Takahiro Sumi}
\affil{Department of Earth and Space Science, Graduate School of Science, The University of Osaka, Toyonaka, Osaka 560-0043, Japan. e-mail: {\tt sumi@ess.sci.osaka-u.ac.jp}}
\author[0000-0001-7862-2070]{David A.~H.~Buckley}
\affiliation{South African Astronomical Observatory, P.O. Box 9, Observatory, 7935, Cape Town, South Africa}
\author[0000-0002-3569-7421]{Alexander S.~Kutyrev}
\affiliation{Department of Astronomy, University of Maryland, College Park, MD 20742, USA}
\affiliation{NASA Goddard Space Flight Center, Greenbelt, MD 20771, USA}
\email{akutyrev@umd.edu}
\author[0000-0002-6510-0681]{Motohide Tamura}
\affiliation{Astrobiology Center, 2-21-1 Osawa, Mitaka-shi, Tokyo 181-8588, Japan}
\affiliation{National Astronomical Observatory, 2-21-1 Osawa, Mitaka-shi, Tokyo 181-8588, Japan}
\affiliation{Department of Astronomy, University of Tokyo, 7-3-1 Hongo, Bunkyo-ku, Tokyo 113-0033, Japan}

%\nocollaboration

\author{David P.~Bennett}
\affiliation{Code 667, NASA Goddard Space Flight Center, Greenbelt, MD 20771, USA}
\affiliation{Department of Astronomy, University of Maryland, College Park, MD 20742, USA}
%\author{Aparna Bhattacharya}
%\affiliation{Code 667, NASA Goddard Space Flight Center, Greenbelt, MD 20771, USA}
%\affiliation{Department of Astronomy, University of Maryland, College Park, MD 20742, USA}
\author{Ian A. Bond}
\affiliation{School of Mathematical and Computational Sciences, Massey University, Auckland 0745, New Zealand}
\author{Giuseppe Cataldo}
\affiliation{NASA Goddard Space Flight Center, Greenbelt, MD 20771, USA}
\author[0000-0002-3774-1270]{Joseph M.~Durbak}
\affiliation{Department of Physics, University of Maryland, College Park, MD 20742, USA}
\affiliation{NASA Goddard Space Flight Center, Greenbelt, MD 20771, USA}
\email{jmdurbak@terpmail.umd.edu}
\author{S. Bradley Cenko}
\affiliation{NASA Goddard Space Flight Center, Greenbelt, MD 20771, USA}
\affiliation{Department of Astronomy, University of Maryland, College Park, MD 20742, USA}
\email{brad.cenko@nasa.gov}
\author{Dale Fixsen}
\affiliation{University of Maryland, College Park, MD 20742, USA}
\affiliation{NASA Goddard Space Flight Center, Greenbelt, MD 20771, USA}
\author{Orion Guiffreda}
\affiliation{Department of Astronomy, University of Maryland, College Park, MD 20742, USA}
\affiliation{NASA Goddard Space Flight Center, Greenbelt, MD 20771, USA}
\email{gregory.mosby@nasa.gov} 
\author{Ryusei Hamada}
\affiliation{Department of Earth and Space Science, Graduate School of Science, The University of Osaka, Toyonaka, Osaka 560-0043, Japan}
\author{Yuki Hirao}
\affiliation{Institute of Astronomy, Graduate School of Science, The University of Tokyo, 2-21-1 Osawa, Mitaka, Tokyo 181-0015, Japan}
\author{Asahi Idei}
\affiliation{Department of Earth and Space Science, Graduate School of Science, The University of Osaka, Toyonaka, Osaka 560-0043, Japan}
\author{Dan Kelly}
\affiliation{Department of Astronomy, University of Maryland, College Park, MD 20742, USA}
\affiliation{NASA Goddard Space Flight Center, Greenbelt, MD 20771, USA}
%\author{Stela Ishitani Silva}
%\affiliation{Department of Physics, The Catholic University of America, Washington, DC 20064, USA}
%\affiliation{Code 667, NASA Goddard Space Flight Center, Greenbelt, MD 20771, USA}
\author{Markus Loose}
\affiliation{Markury Scientific, Inc. 518 Oakhampton Street, Thousand Oaks, CA 91361, USA}
\email{markus.loose@markury-scientific.com}
\author{Gennadiy N. Lotkin}
\affiliation{NASA Goddard Space Flight Center, Observational Cosmology Laboratory, Greenbelt, MD 20771, USA}
\affiliation{Beacon Systems, Inc., Washington, DC 20024, USA}
\email{gennadiy.lotkin@beacongov.com}
\author{Eric I. Lyness}
\affiliation{NASA Goddard Space Flight Center, Planetary Environments Laboratory, Greenbelt, MD 20771, USA}
\affiliation{Microtel LLC, Greenbelt, MD 20770, USA}
\email{eric.i.lyness@nasa.gov}
\author{Stephen Maher}
\affiliation{NASA Goddard Space Flight Center, Observational Cosmology Laboratory, Greenbelt, MD 20771, USA}
\affiliation{Science Systems and Applications, Inc., Lanham, MD 20706, USA}
\email{stephen.f.maher@nasa.gov}
\author{Shuma Makida}
\affiliation{Department of Earth and Space Science, Graduate School of Science, The University of Osaka, Toyonaka, Osaka 560-0043, Japan}
\email{makida@iral.ess.sci.osaka-u.ac.jp}
\author[0000-0001-7936-0898]{Noriyuki Matsunaga}
\affiliation{Department of Astronomy, Graduate School of Science, The University of Tokyo, 7-3-1 Hongo, Bunkyo-ku, Tokyo 113-0033, Japan}
\email{matsunaga@astron.s.u-tokyo.ac.jp}
\author{Shota Miyazaki}
\affiliation{Institute of Space and Astronautical Science, Japan Aerospace Exploration Agency, 3-1-1 Yoshinodai, Chuo, Sagamihara, Kanagawa 252-5210, Japan}
\author{Gregory Mosby}
\affiliation{NASA Goddard Space Flight Center, Greenbelt, MD 20771, USA}
\author{Samuel H. Moseley}
\affiliation{NASA Goddard Space Flight Center, Observational Cosmology Laboratory, Greenbelt, MD 20771, USA}
\affiliation{University of Maryland, College Park, MD 20742, USA}
\author{Tutumi Nagai}
\affiliation{Department of Earth and Space Science, Graduate School of Science, The University of Osaka, Toyonaka, Osaka 560-0043, Japan}
\author{Togo Nagano}
\affiliation{Department of Earth and Space Science, Graduate School of Science, The University of Osaka, Toyonaka, Osaka 560-0043, Japan}
\author{Seiya Nakayama}
\affiliation{Department of Earth and Space Science, Graduate School of Science, The University of Osaka, Toyonaka, Osaka 560-0043, Japan}
\author{Mayu Nishio}
\affiliation{Department of Earth and Space Science, Graduate School of Science, The University of Osaka, Toyonaka, Osaka 560-0043, Japan}
\author{Kansuke Nunota}
\affiliation{Department of Earth and Space Science, Graduate School of Science, The University of Osaka, Toyonaka, Osaka 560-0043, Japan}
\author{Ryo Ogawa}
\affiliation{Department of Earth and Space Science, Graduate School of Science, The University of Osaka, Toyonaka, Osaka 560-0043, Japan}
\author{Ryunosuke Oishi}
\affiliation{Department of Earth and Space Science, Graduate School of Science, The University of Osaka, Toyonaka, Osaka 560-0043, Japan}
\author{Yui Okumoto}
\affiliation{Department of Earth and Space Science, Graduate School of Science, The University of Osaka, Toyonaka, Osaka 560-0043, Japan}
%\author{Greg Olmschenk}
%\affiliation{Code 667, NASA Goddard Space Flight Center, Greenbelt, MD 20771, USA}
%\author{Cl\'ement Ranc}
%\affiliation{Sorbonne Universit\'e, CNRS, UMR 7095, Institut d'Astrophysique de Paris, 98 bis bd Arago, 75014 Paris, France}
\author[0000-0001-5069-319X]{Nicholas J. Rattenbury}
\affiliation{Department of Physics, University of Auckland, Private Bag 92019, Auckland, New Zealand}
\author[0000-0002-1228-4122]{Yuki K. Satoh}
\affiliation{College of Science and Engineering, Kanto Gakuin University, Yokohama, Kanagawa 236-8501, Japan}
\author{Elmer H. Sharp}
\affiliation{NASA Goddard Space Flight Center, Observational Cosmology Laboratory, Greenbelt, MD 20771, USA}
\affiliation{Center for Space Sciences and Technology, University of Maryland, Baltimore County, Baltimore, MD 21250, USA}
%\author{Takahiro Sumi}
%\affiliation{Department of Earth and Space Science, Graduate School of Science, The University of Osaka, Toyonaka, Osaka 560-0043, Japan}
\author{Daisuke Suzuki}
\affiliation{Department of Earth and Space Science, Graduate School of Science, The University of Osaka, Toyonaka, Osaka 560-0043, Japan}
\author{Takuto Tamaoki}
\affiliation{Department of Earth and Space Science, Graduate School of Science, The University of Osaka, Toyonaka, Osaka 560-0043, Japan}
\email{tamaoki@iral.ess.sci.osaka-u.ac.jp}
%\author{Sean K. Terry}
%\affiliation{Code 667, NASA Goddard Space Flight Center, Greenbelt, MD 20771, USA}
%\affiliation{Department of Astronomy, University of Maryland, College Park, MD 20742, USA}
\author{Eleonora Troja}
\affiliation{Department of Physics, University of Rome Tor Vergata, Via della Ricerca Scientifica 1, 00133 Rome, Italy}
\affiliation{NASA Goddard Space Flight Center, Greenbelt, MD 20771, USA}
\email{nora@roma2.infn.it}
%\author{Aikaterini Vandorou}
%\affiliation{Code 667, NASA Goddard Space Flight Center, Greenbelt, MD 20771, USA}
%\affiliation{Department of Astronomy, University of Maryland, College Park, MD 20742, USA}
\author[0000-0002-2340-8303]{Sarah V. White}
\affiliation{South African Astronomical Observatory, P.O. Box 9, Observatory, 7935, Cape Town, South Africa}
\author{Hibiki Yama}
\affiliation{Department of Earth and Space Science, Graduate School of Science, The University of Osaka, Toyonaka, Osaka 560-0043, Japan}
\collaboration{(PRIME collaboration)}


\begin{abstract}
We present the description of the instruments and the first results of 
the PRime-focus Infrared Microlensing Experiment (PRIME).
PRIME is the first dedicated near-infrared (NIR) microlensing survey telescope located 
 at the South African Astronomical Observatory (SAAO) in Sutherland, South Africa.
Among its class, it offers one of the widest fields of view in the NIR regime.
PRIME's main goals are
(1) To study planetary formation by measuring the frequency and mass function of planets. 
In particular, we compare results from the central Galactic bulge (GB), 
accessible only in the NIR by PRIME, with those from the outer GB by optical surveys.
(2) To conduct concurrent observations with NASA’s Nancy Grace Roman Space telescope.
Due to the different lines of sight between the ground and space, 
we detect slight variations in light curves,  known as ``Space-based parallax."
This effect allows us to measure the mass of lens systems and their distance from the Earth.
It is the only method to measure the mass of the free-floating planets down to Earth-mass.
We began the GB survey in February 2024 and analyzed images through June 1, 2025, 
identifying 486 microlensing candidates and over a thousand variable stars, 
including Mira variables, which are useful to study the Galactic structure.
We issue real-time alerts for follow-up observations, supporting exoplanet searches,
 and the chemical evolution studies in the GB.
During the off-bulge season, we conduct an all-sky grid survey and Target of Opportunity (ToO) 
observations of transients, including gravitational wave events, $\gamma$-ray bursts, 
and other science.
\end{abstract}




\section{Introduction} \label{sec:intro}

The gravitational microlensing observations toward the Galactic bulge (GB) enable 
us to study the structure, kinematics and dynamics of the Galaxy 
(\citealt{pac91}, \citealt{gri91}, \citealt{sumiPenny2016}, \citealt{Mroz19}, \citealt{Nunota2024b}),
and the measurement of the stellar and free-floating planets (FFP) mass functions (MFs) 
\citep{pac91,sumi2011, Wegg2017, Mroz17,Mroz19, Mroz20a, Koshimoto2023,sumi2023} 
and exoplanet searches \citep{mao1991,gaudi-ogle109,bennett-ogle109, suzuki2016, kos21b,Nunota2024a}.

To date, several tens of thousand microlensing events have been detected toward the GB 
by various microlensing survey groups:
OGLE (\citealt{uda94, uda00, woz01, uda03,sumi2006}), 
MOA (\citealt{bon01}, \citealt{sumi03}), MACHO (\citealt{alc97}, \citealt{alc00b}),
EROS (\citealt{afo03}, \citealt{ham06}), WiSE  \citep{wise_survey} and KMTNet \citep{zang23}.
Thousands of additional microlensing events are expected to be detected in the coming years 
by MOA-II, OGLE-IV, and KMTNet, which are currently in operation.
%MOA-II\footnotemark\footnotetext{\tt http://www.massey.ac.nz/\~{}iabond/alert/alert.html}, 
%OGLE-IV\footnotemark\footnotetext{\tt http://www.astrouw.edu.pl/\~{}ogle/ogle4/ews/ews.html},
%and
%WiSE\footnotemark\footnotetext{\tt http://wise-obs.tau.ac.il/\~{}wingspan/} \citep{wise_survey}
%surveys, which are currently in operation. These surveys will soon be joined by the 
%KMTNet survey \citep{kmtnet}.

The magnification of a microlensing event is described by (\citealt{pac86})

\begin{equation}
  \label{eq:amp-u}
  A(u)= \frac{u^2+2}{u\sqrt{u^2+4}},
\end{equation}

\noindent
where $u$ is the projected separation of the source and lens 
in units of the angular Einstein radius  $\theta_{\rm E}$ given by
 %  $\theta_{\rm E}=\mu_{\rm rel}/t_{\rm E}=9.2 \mu$
 \begin{equation}
  \label{eq:thetaE}
\theta_{\rm E} =t_{\rm E}  \mu_{\rm rel} = \sqrt{\kappa M \pi_{\rm rel}}.
\end{equation}
Here, $\kappa =4G/(c^2 {\rm au})= 8.144 {\rm mas}/M_\odot$,
the lens-source relative proper motion is $\mu_{\rm rel}$, 
the lens-source relative parallax is
$\pi_{\rm rel}=\pi_{\rm \ell}^{-1}- \pi_{\rm s}^{-1}=1$ au$(D_{\rm \ell}^{-1}-D_{\rm s}^{-1})$.
The lens mass $M$, the distance $D_{\rm \ell}$ to the lens and the relative proper motion $\mu_{\rm rel}$ 
are degenerate in the Einstein timescale,  $t_{\rm E}= \theta_{\rm E}/\mu_{\rm rel} $,
where $D_{\rm s}$ is the distance to the source star.
This implies that the mass function of the lens population must be determined statistically, 
based on an assumed model of the stellar population density and velocity distribution in the Galaxy.
In other words, the galactic characteristics --- the population density and kinematics --- can be extracted from the observed 
$t_{\rm E}$ distribution.

However, all of these existing surveys operate at optical wavelengths and are 
therefore limited to the outer regions of the Galactic bulge (GB), 
at higher Galactic latitudes, where dust extinction is relatively low.
To observe the central region of the GB—i.e., at lower Galactic latitudes, 
where stellar density and dust extinction are higher—a near-infrared (NIR) telescope is required.
Observations in the central GB offer several advantages:
(1) the microlensing event rate is higher.
(2) We can measure the exoplanet occurrence rate in these dense stellar fields and 
compare it to that in the outer regions as measured by optical surveys. 
This allows us to investigate environmental effects on planet formation in the Galactic scale.
(3) We can study the structure of the central GB, complementing the current optical surveys of the outer GB.

So far, two NIR survey campaigns have been conducted toward the central GB.
The 3.8 m UKIRT telescope on Mauna Kea, Hawaii, observed the GB in the $H$ and $K_{s}$ bands
with a cadence of 0.3-3 observations per night \citep{Shvartzvald2018}.
This was the first NIR microlensing campaign. 
They found about one thousand microlensing events in the central GB region, 
including 24 anomalous events, one of which was due to an exoplanet detection \citep{Han2025}.
However, the number of detected events was limited because
 it was not a long-term, dedicated survey.
The VISTA Variables in the Via Lactea (VVV) survey conducted 
wide-field, multi-epoch observations of the GB and part of the disk in the $K_{s}$ band, 
reaching a depth of 18 mag over a total area of 560 square degrees \citep{Minniti2010}.
VVV detected approximately one million variable objects through hundreds of observations conducted between 2010 and 2019.
The Red Clump (RC) giants and RR Lyrae observed in the VVV survey are used to model the three-dimensional structure of the Galactic bar and bulge \citep{Gonzalez2011, Gran2016, Simion2017}.
Although the survey was not specifically designed to detect microlensing events, 
it nevertheless identified 1,959 microlensing events toward the GB.
However, its cadence was not high enough to detect exoplanets \citep{Navarro2020,Husseiniova2021}.

The WINTER 1-m telescope\footnote{\tt https://www.daniellefrostig.com/research/winter} 
uses InGaAs sensors, which are less sensitive than HgCdTe detectors, 
and is located in the northern hemisphere. 
Therefore, it is not well suited for the Bulge survey.
The DREAMS telescope\footnote{\tt https://dreams.anu.edu.au} 
conducts an all-sky monitoring survey in the Southern Hemisphere. 
However, it has an even smaller aperture (0.5 m) and 
employs less sensitive InGaAs sensors.

NASA's Astrophysics Division's next flagship mission the Nancy Grace Roman Space Telescope
 is a 2.4-meter NIR and optical telescope, equipped with a wide-field (0.28 deg$^2$)  NIR camera and optical coronagraph to be launched in October 2026.
Roman is planning to conduct the Roman Galactic Bulge Time Domain Survey (GBTDS) in the NIR and 
is expected to detect approximately $\sim$27,000 microlensing events
and $\sim$1,400 exoplanets, including 200 with masses less than 3 Earth masses 
\citep{Johnson2020}, as well as around 1,000 free-floating planets (FFPs) \citep{sumi2023}.
The candidate survey fields include five located in the outer GB and one at the Galactic center, 
including the region around Sgr A* \cite{Zasowski2025}.

Microlensing surveys provide valuable datasets for the study of pulsating stars as a by-product, as well demonstrated, for example, by OGLE (\citealt{Soszynski2017, Iwanek2022}). PRIME is also expected to discover a large number of pulsating stars, such as Mira variables, RR Lyrae variables, and Cepheids. Its multi-band NIR time-series data will be particularly important for extending catalogs of variable stars in the highly reddened regions toward the GB. At optical wavelengths, high interstellar extinction has prevented the detection of even bright Miras within $\pm$1 degree of the Galactic plane by OGLE \citep{Iwanek2022} or Gaia \citep{Mowlavi2018,Lebzelter2023}. In contrast, the VVV survey has collected long-term $K_s$-band data over a large area covering the GB and southern Galactic disk, discovering many variable stars across the Galactic plane \citep{Sanders2022, Albarracin2025}. However, a significant fraction of Miras at the distance of the GB are too bright and become saturated in the $K_s$-band images of VVV. PRIME can observe such bright Miras, especially at shorter NIR wavelengths where interstellar extinction is more significant. By combining time-series data in multiple bands ($ZYJH$), it becomes possible to detect variable stars toward the GB across a wide range of interstellar extinction. The pulsating stars detected in this way will be valuable tracers for studies of the Galactic structure.

The Prime-focus Infrared Microlensing Experiment (PRIME) is the first dedicated 
wide-field NIR microlensing survey telescope, located at the Sutherland Observatory in South Africa.
It  has been built within the frameworks of the collaboration of NASA and JAXA on the Roman Space Telescope. 
In this paper, we present the design and current status of the PRIME telescope, 
as well as the survey plans for both the GB and off-bulge seasons.
We also report preliminary results from the first observing season.
An overview of the PRIME project is provided in Section~\ref{sec:PRIME}.
Details of the telescope and camera are described in Section~\ref{sec:Telescope}.
The survey strategy is presented in Section~\ref{sec:observation}, 
and the data reduction process and initial results are summarized in Section~\ref{sec:Data_reduction}.
The all-sky grid observations are described in Section~\ref{sec:ALL-SKY}.
Finally, discussion and conclusions are given in Section~\ref{sec:discussionAndSummary}.

% Appendix~\ref{sec-append}.

%\begin{enumerate}
%\item improved citations for third party data repositories and software,
%\item easier construction of matrix figures consisting of multiple 
%encapsulated postscript (EPS) or portable document format (PDF) files,
%\item figure set mark up for large collections of similar figures,
%\end{enumerate}

\section{PRIME project} \label{sec:PRIME}
The PRIME project is a collaboration between the University of Osaka, the Astrobiology Center (ABC), NASA, the University of Maryland (UMd), and the South African Astronomical Observatory (SAAO). It represents one of Japan's contributions to NASA's Roman Space Telescope mission, conducted through the JAXA-led Roman-J project. 
The University Osaka and the ABC led the construction of the telescope and its dome, while the University of Osaka and the UMd jointly developed the NIR camera, which incorporates four NIR detectors provided by NASA’s Roman Project. The SAAO played a central role in the construction of the observatory building.

The primary goal of PRIME is to investigate the formation mechanisms of exoplanets by conducting the first dedicated NIR gravitational microlensing exoplanet survey toward the inner GB. 
This includes both precursor and concurrent observations with the Roman Space Telescope.
Half of the telescope time will be allocated to a time-domain survey of the GB whenever it is observable. The remaining 50\% will be shared among the partner institutions for their own scientific interests, such as studies of variable stars, 
transiting exoplanet searches, and ToO observations of gravitational wave events, gamma-ray bursts, and other transient phenomena.

The wide field of view (FOV) in the NIR provided by PRIME enables survey observations of the inner GB at low Galactic latitudes, including the region around Sgr A, where the stellar density is higher than in the outer GB at higher latitudes, which are typically targeted by conventional optical surveys.
Taking advantage of these capabilities, the primary objectives of PRIME compared to conventional optical surveys are as follows:
\begin{itemize}
\item[$(1)$]
We will increase the statistics of microlensing events, including bound exoplanets and FFPs, due to the higher event rate in the inner Galactic bulge.

\item[$(2)$]
We will measure exoplanet occurrence rates in the high stellar density fields of the inner GB and compare them with those in the outer GB observed by conventional optical surveys. This allows us to investigate the environmental dependence of planet formation.

\item[$(3)$]
We will create maps of the microlensing event rate and optical depth in the inner GB, which can be used to constrain Galactic structure --- the population density and kinematics ---  models . 
These results will complement existing optical surveys of the outer GB.
These maps will also inform the selection of target fields for the Roman Space Telescope.

\item[$(4)$]
After the launch of Roman in late 2026, PRIME will conduct simultaneous observations of the same fields as Roman. The slight difference in the line of sight between the ground and space will produce variations in the light curves—an effect known as space-based microlens parallax \citep{Udalski2015}. This enables direct measurement of the lens system's mass and distance from Earth. It is the only method currently available to measure the mass of FFPs \citep{sumi2023, Koshimoto2023}.

\item[$(5)$]
We will detect variable stars, including Mira variables located in or behind the GB and disk, which are too faint for 2MASS and too bright for VVV. These stars will be valuable tracers for studying the structure of the Galaxy.
\end{itemize}

\section{PRIME Telescope} \label{sec:Telescope}
\subsection{Telescope} \label{sec:telescope}

\begin{figure}
\begin{center}
\includegraphics[scale=0.5,keepaspectratio]{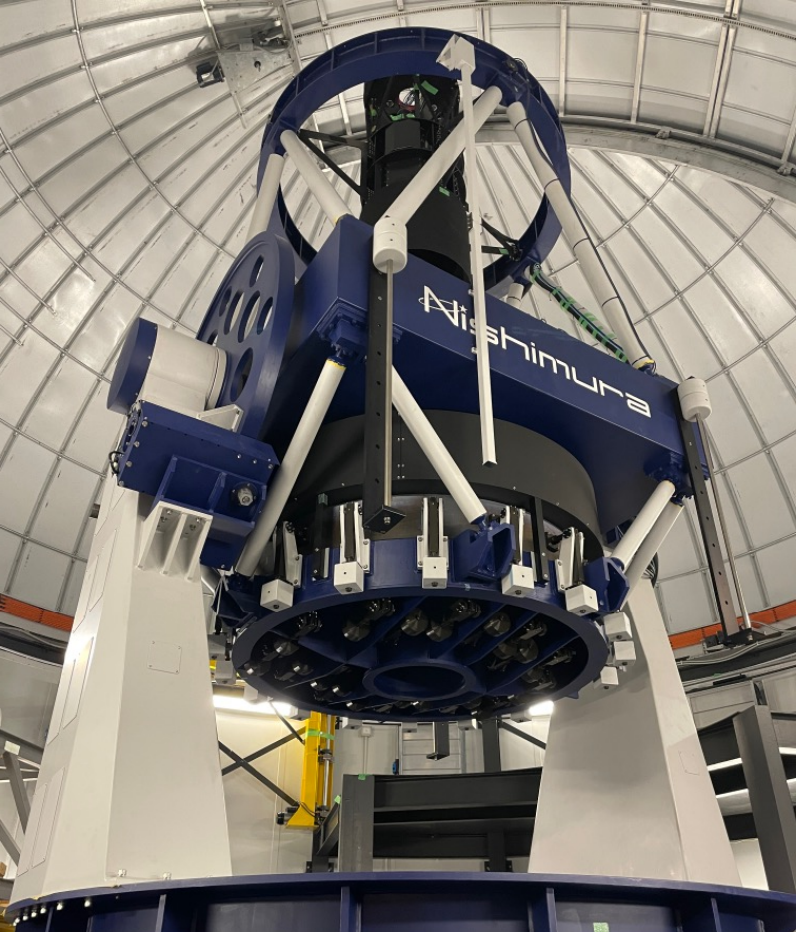}
\caption{
  \label{fig:PRIME_telescope}
PRIME 1.8m telescope.}
\end{center}
\end{figure}

\begin{figure}
\begin{center}
\includegraphics[scale=0.5,keepaspectratio]{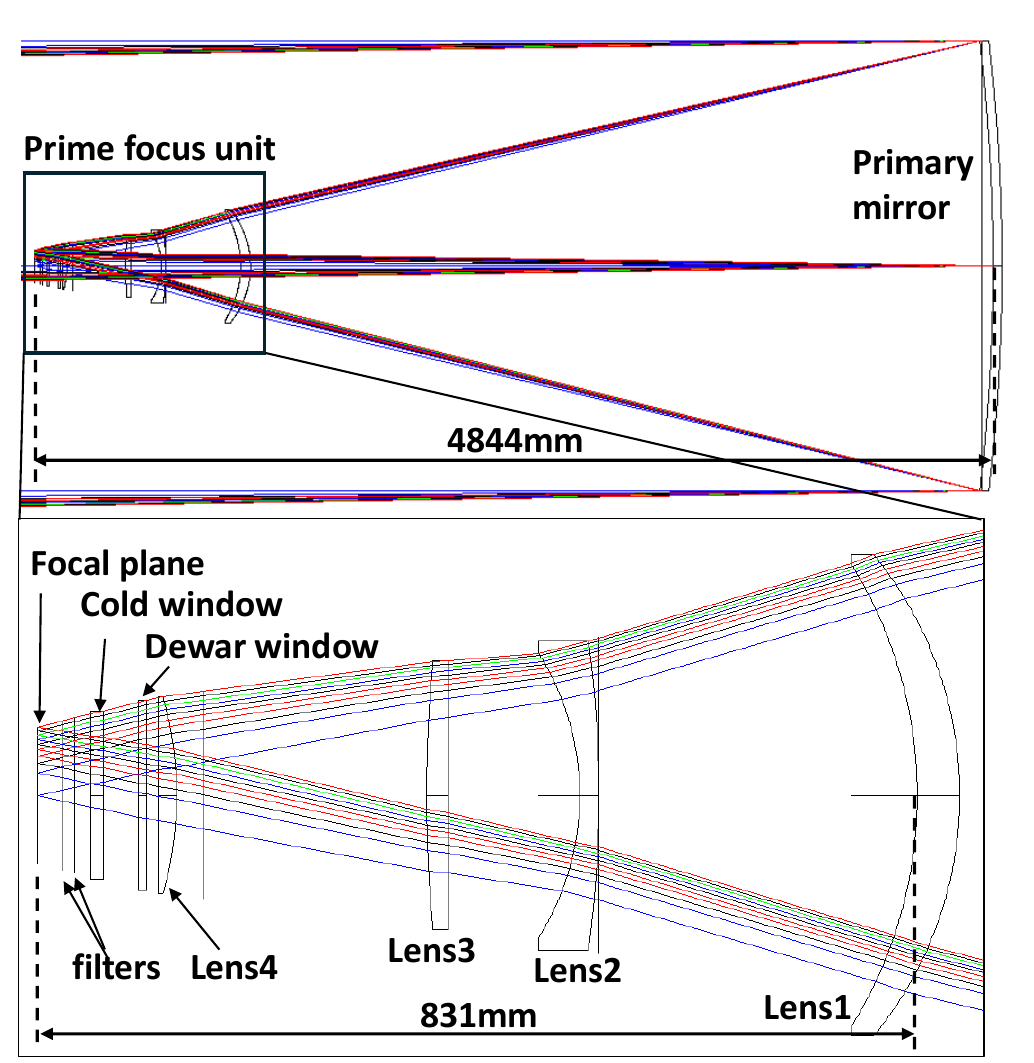}
\caption{
  \label{fig:Optical_design}
The optical design of the PRIME telescope with ray tracing.
The top panel presents the entire optical system.
The rays traveling to the detector along paths near the optical axis 
are blocked by the prime-focus camera in reality.
These are shown here for illustrative purposes only.
The bottom panel provides a close-up view of the prime-focus lens unit.}
\end{center}
\end{figure}

The PRIME telescope is a 1.8-m aperture NIR prime-focus telescope
(Figure \ref{fig:PRIME_telescope} and see Table \ref{tbl:PRIMEtelescope} for the telescope properties).
Its primary mirror is a 1.8-m paraboloid with a focal length of 3792 mm (f/2.11), 
made of AstroSitall. The mirror is coated with aluminum (Al) and silicon dioxide (SiO$_2$) for durability, providing a reflectivity of 92-95\% in the 1000-1,800 nm wavelength range.
The wide FOV is achieved using a prime focus corrector unit (PFU) consisting of four spherical fused silica lens elements. The first element has an edge diameter of 465.09 mm. The second element, with a diameter of 305.44 mm, is mounted on a tip/tilt mechanism that can be remotely adjusted using three motors for fine optical alignment.
Elements 1, 3 and 4 are fixed. The entire PFU can be moved to adjust the focus.
See Table~\label{tbl:lenses} for details of the optical specifications of the corrector lenses.
The combined focal length of the primary mirror and the corrector unit is 4125 mm (f/2.29). Figure~\ref{fig:Optical_design} shows the optical design of the telescope with ray tracing.

\begin{figure}
\begin{center}
\includegraphics[scale=0.7,keepaspectratio]{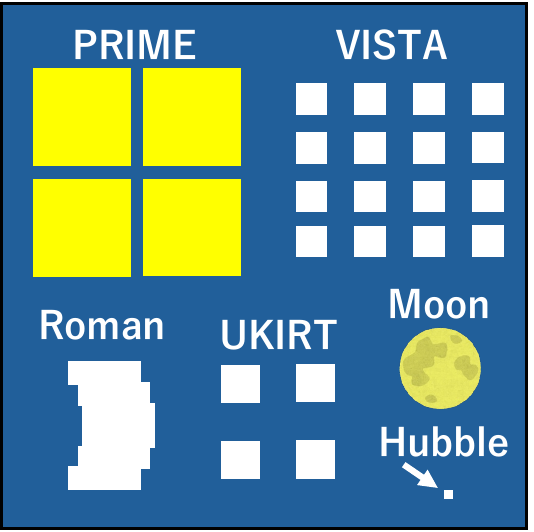}
\caption{
  \label{fig:FOV}
Field of view (FOV) of the PRIME telescope ($1.45$ deg$^2$, with gaps) 
compared to those of VISTA, Roman ($0.28$ deg$^2$), UKIRT, Hubble, and the full Moon.
}
\end{center}
\end{figure}

The 80\% encircled energy diameter (EED) is designed to be 14 $\mu$m (1.4 pixels, 0.7 arcsec) across the entire FOV, 
which is sufficient under the typical seeing conditions of approximately 1.4 arcseconds at Sutherland 
in optical \citep{Catala2013}.
Here, the observed full width at half maximum (FWHM) of the point spread function (PSF) 
in our $H$-band images is $\sim$ 1.3 arcsec, as shown in Section~\ref{sec:Data_reduction}. 
The FOV is $1.14^{\circ} \times 1.14^{\circ} = 1.29$ deg$^2$ excluding detector gaps, 
and $1.21^{\circ} \times 1.21^{\circ} = 1.45$ deg$^2$ including them. 
Figure~\ref{fig:FOV} shows the FOV of the PRIME telescope in comparison with other telescopes 
and the size of the full Moon.
The optical system is optimized for NIR observations, with a bandwidth of 830-1800 nm, 
covering the range from the $Z$-band to the $H$-band.

The telescope has been installed on top of the pier, 6m above the ground, to improve the seeing.

%H4RG-10:   40.96mm (4096pix) + gap 4.91mm (491pix) = 86.83mm
%モデルのH4RGの検出面の一辺の長さが40.88mmなので、40.96mmと0.08mmの誤差があり
%ギャップも一つは4.94mm,もう一つが4.67mmで完全な正方形ではないです。
%伊藤さんの光学モデルでは検出面の１辺を86.83mm（491pix）で計算しているので、この値を使ったらどうでしょうか？
%One side of FOV  with the gap of 491pix: (8192pix + 491pix)x 0.5”/pix = 4341.5” = 1.206deg.   (FOV=1.454 deg^2)
%Without gap:                                             (8192pix               )x 0.5”/pix = 4096” = 1.138deg. (FOV= 1.2945 deg^2)

%------------------------Table 1.---------------------------------
\begin{deluxetable}{lrrrrrc}
\tabletypesize{\scriptsize}
%\rotate
\tablecaption{Optical specification of the PRIME telescope
\label{tbl:PRIMEtelescope}
}
\tablewidth{0pt}
\tablehead{
\colhead{Item} &
\colhead{value} 
%\colhead{} & 
%\colhead{$(\rm day)$} & 
%\colhead{} & 
%\colhead{} 
}
\startdata
Primary material        &  AstroSitall CO-115M            \\
Primary diameter       & 1800mm    \\
Conic Constant         &  -1              \\
Primary focal length   & 3792mm (f/2.11)            \\ 
Mirror Coating             & Al+SiO$_2$        \\ 
Focus                          &  Prime focus          \\
Combined focal length  & 4125mm (f/2.29)             \\
Pixel scale               &   0.5 "/pix on 10$\mu$m pixel   \\
Number of Pixel       &   4096$\times$4096 $\times$ 4 pixels  \\
Field of view w/o gap     &   1.14$^{\circ}$  $\times$ 1.14$^{\circ}$ =1.29 deg.$^2$         \\      
Field of view w/ gap       &   1.21$^{\circ}$  $\times$ 1.21$^{\circ}$ =1.45 deg.$^2$         \\ 
Bandwidth               &  830nm-1800nm             \\
 \enddata
 %\tablenotetext{a}{Likely Brown dwarf lens.}
%\table comments{
%}
\end{deluxetable}
%----------------------------------------------------------------

%------------------------Table 2.---------------------------------
\begin{deluxetable}{lrrrrrc}
\tabletypesize{\scriptsize}
%\rotate
\tablecaption{Optical specifications of the corrector lenses
\label{tbl:lenses}
}
\tablewidth{0pt}
\tablehead{
\colhead{Element} &
%\colhead{material} &
%\colhead{Conic Constant} &
\colhead{Diameter} &
\colhead{ROC1\tablenotemark{$a$}} &
\colhead{ROC2\tablenotemark{$a$}} &
\colhead{Thickness\tablenotemark{$b$}} &
\colhead{Distance\tablenotemark{$c$}} \\
\colhead{} &
\colhead{(mm)} &
\colhead{(mm)} &
\colhead{(mm)} &
\colhead{(mm)} &
\colhead{(mm)} 
%\colhead{value} \\
%\colhead{} & 
%\colhead{$(\rm day)$} & 
%\colhead{} & 
%\colhead{} 
}
\startdata
%Primary           & 1800   &  -     & - & - & 2968 \\ 
%space                      &                     &                  &             & 2968\\ 
Lens1                      & 465.09   &      357.49  &  411.92  & 44.54 & 2968\\
%space                      &                     &                  &             & 289.35\\ 
Lens2                      & 305.44   &    1071.92  &  254.57  & 17.72 & 289.35\\ 
%space                      &                     &                  &             & 129.67\\ 
Lens3                      & 270.01   &    plano    &  1364.84  & 19.97 & 129.67\\ 
%space                      &                     &                  &             & 243.54\\ 
Lens4                      & 199.95   &    348.17  &  plano  & 18.99 & 243.54\\ 
 \enddata\
  \tablenotetext{a}{Radius of curvature of Surface 1 (closer to the primary) and 2.}
  \tablenotetext{b}{Axial thickness.}
 \tablenotetext{c}{Distance from the surface of the previous element along the optical axis. 
 (I.e, distance from the primary mirror for Lens1)}
 %\tablenotetext{a}{Likely Brown dwarf lens.}
%\table comments{
%}
\end{deluxetable}
%----------------------------------------------------------------

%----------------------------- FIG. 1 -------------------------------------
%--------------------------------------------------------------------------

%----------------------------- FIG. 2 -------------------------------------
%--------------------------------------------------------------------------

%----------------------------- FIG. 3 -------------------------------------
%--------------------------------------------------------------------------

The telescope is located at the Sutherland Observatory in South Africa, at an altitude of 1780 m.
The observatory is operated by the SAAO.
The observatory building was constructed by SAAO with funding from the National Research Foundation (NRF).
Although the construction of the building and the installation of the telescope were delayed due to the COVID-19 pandemic, both were completed in July 2022.
Optical alignment was carried out using the $H$-band test camera, YAMA-cam, and the expected optical performance was confirmed \citep{Yama2023}.

%We use the microlensing sample selected from the MOA-II high cadence photometric survey 
%toward the Galactic Bulge in the 2006-2014 seasons \citep[][hereafter \citetalias{Koshimoto2023}]{Koshimoto2023}.
%MOA-II uses the 1.8-m MOA-II telescope which has 
%a 2.18 deg$^2$ field of view (FOV) and which is
%ocated at the Mt.\ John University
%Observatory, New Zealand\footnote{\url{https://www.massey.ac.nz/~iabond/moa/alerts/}}.
%We briefly summarize here.

%%%%%%%%%%%%%%%%%%%%%%%%%%%%%%%%%%%
\subsection{PRIME-Cam} \label{sec:RPIME-cam}
The main instrument of the PRIME telescope is a wide-field NIR prime-focus camera, 
called PRIME-Cam \citep{Kutyrev2023,Durbak2024}.
The University of Maryland developed the camera at NASA’s Goddard Space Flight Center (GSFC).
It employs four Teledyne H4RG-10 detectors, each with 4096$\times$4096 pixels, 10$\mu$m pixel size, and a 2.5$\mu$m cutoff wavelength—specifications identical to those used in the Roman Space Telescope \citep{Mosby2020}.
Each detector measures 40.96 mm$\times$40.96 mm, and the total focal plane area is 86.83 mm$\times$86.83 mm, including a 4.91 mm gap between detectors (equivalent to 491 pixels).

The four detectors are read out by four ACADIA electronics located inside the dewar and two MACIE electronics located outside the dewar, with a readout noise of 5 ADU per read \citep{Loose2018}.
These detectors and readout electronics are on loan from the NASA's Roman project 
as an in-kind contribution to PRIME. 
ACADIA performs non-destructive readouts of all pixels using the Sample Up The Ramp (SUTR) method, acquiring one frame every 2.86 seconds (Kutyrev et al. in preparation).

\subsection{Filters} \label{sec:filters}
The dewar window initially consisted of 9-mm-thick fused silica and transmitted light in the 830-1800 nm range.
However, it was replaced with a 6-mm-thick sapphire window in 2025 to prevent condensation.
A cold window made of BK7 is installed on the front surface of the cold box to 
block the light outside of the 800-1900 nm science range thus effectively blocking thermal radiation.
Specifically, it blocks light from 1950-2200 nm with an optical density (OD) greater than 3.5, 
and from 2200-3000 nm with an OD greater than 5, in combination with bandpass filters.

Each of the two filter wheels has four slots, providing a total of eight available positions.

\begin{figure}
\begin{center}
\includegraphics[scale=0.6,keepaspectratio]{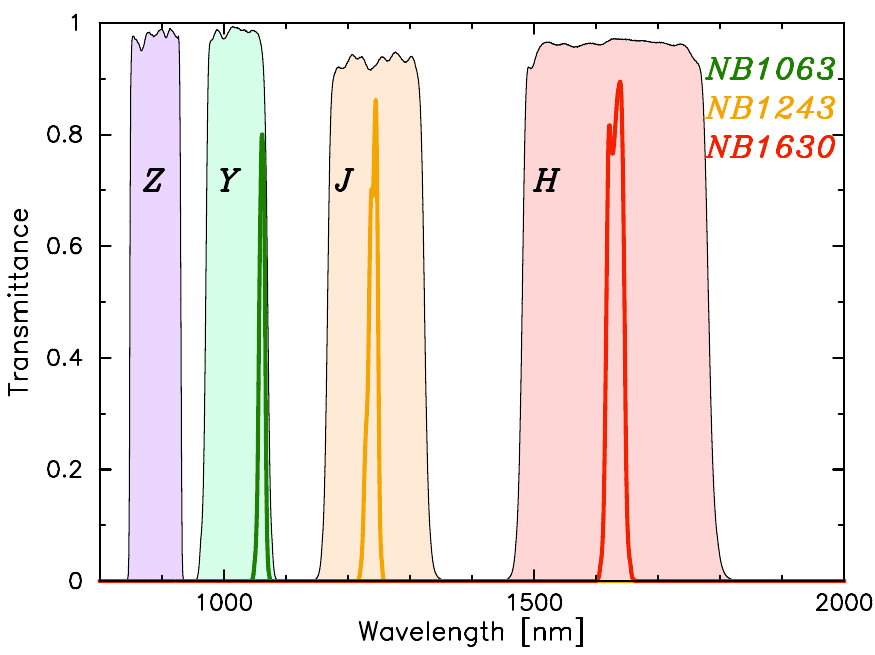}
\caption{
  \label{fig:filter}
Throughput of the filters.
The broad bands Z, Y, J, and H are shown as purple, green, yellow, and 
red shaded regions, respectively.
The narrow bands NB1063, NB1243, and NB1630 are represented by 
thick green, yellow, and red lines, respectively.
}
\end{center}
\end{figure}

Wheel 1 includes a $Z$-band filter, a narrow-band (NB) filter, a blocked position, and an open position.
Wheel 2 contains standard $Y$-, $J$-, and $H$-band filters, along with an open position.
To use a specific filter in one wheel, a blank must be selected in the other wheel.
The blocked position is used for taking dark frames.
The NB filter contains three narrow-band regions printed on a single substrate, 
targeting low-OH sky windows in the $Y$, $J$, and $H$ bands. These are:
(i) 1.063$\mu$m (NB1063), which is useful for characterizing the H4RG-10 detectors due to its low background;
(ii) 1.243$\mu$m (NB1243), corresponding to H$\alpha$ at $z=0.9$ and [O III] at $z=1.5$;
(iii) 1.630$\mu$m (NB1630), corresponding to H$\alpha$ at $z=1.5$.
Each narrow band can be selected by combining the NB filter in wheel 2 with one of the three broad-band filters ($Y$, $J$, $H$) in wheel 1.
All broad and narrow-band filters block out-of-band light with an OD greater than 3.5, to prevent ghosting.
The throughput of the filters are shown in Figure \ref{fig:filter} and Table \ref{tbl:filters}.
The limiting magnitudes (for a 100-second exposure at 5$\sigma$) are 
about 21.0, 20.1, 19.4, and 18.5 mag (Vega) in the $Z$, $Y$, $J$, and $H$ bands, respectively.

The typical sky background levels are approximately $\sim$300, $\sim$900, $\sim$1,000, 
and $\sim$3,000 ADU per 2.86-second exposure in the $Z$, $Y$, $J$, and $H$ bands, respectively,
varying by a factor of a few.
Given that the potential well of the detectors is about 100,000 electron, 
i.e, 60,000 ADU with a gain of $\sim$1.8[e/ADU], 
a single exposure in the $H$-band is limited to four frames (frame 0-3), 
corresponding to $2.86 \times 3 = 8.58$ seconds.
A 51-second total exposure, composed of six such exposures, requires 140 seconds including overhead.
Thus, observing 35 fields takes approximately 82 minutes.

%----------------------------- FIG. 4 -------------------------------------
%--------------------------------------------------------------------------

%------------------------Table 3.---------------------------------
\begin{deluxetable}{lrrrrrc}
\tabletypesize{\scriptsize}
%\rotate
\tablecaption{Filters of the \ telescope
\label{tbl:filters}
}
\tablewidth{0pt}
\tablehead{
\colhead{band} &
\colhead{bandpass} &
\colhead{limiting mag\tablenotemark{$a$}} \\
\colhead{} &
\colhead{(nm)} &
\colhead{(mag)} 
}
\startdata
$Z$      &  850-930    & 21.0 \\
$Y$       & 970-1075  & 20.1\\
$J$       &  1170-1330  & 19.4\\
$H$      & 1490-1780    & 18.5 \\ 
NB1063($Y$)   &   1063.0 $\pm$ 10   & - \\
NB1243($J$)    &   1243.3   $\pm$ 17.7  & - \\
NB1630($H$)   &   1629.6   $\pm$ 23.6 & - \\
 \enddata
 \tablenotetext{a}{5$\sigma$ and 100 sec exposure.}
%\table comments{
%}
\end{deluxetable}
%--------------------------------------------------------------------------

%==================================================================================

%\section{FSPL events}
\section{OBSERVATION}
\label{sec:observation}
\begin{figure}
\begin{center}
\includegraphics[scale=0.35,keepaspectratio]{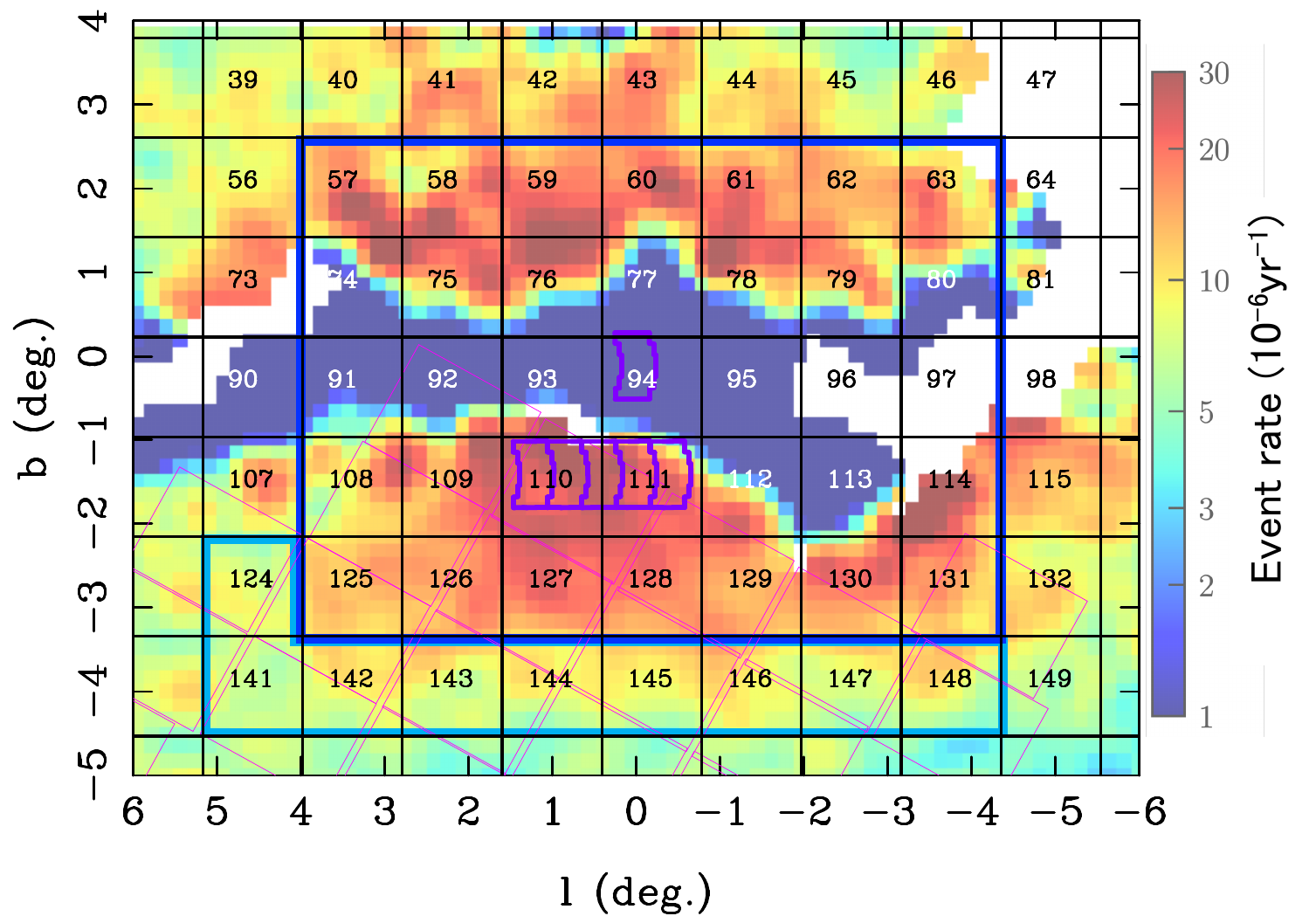}
\caption{
  \label{fig:fields}
Observational fields of PRIME toward the GB.
The color map shows the microlensing event rate based on OGLE data \citep{Mroz19}.
Black square grids and field numbers indicate the PRIME fields, labeled as GB\#.
The blue squares represent the 35 primary GB fields.
Cyan squares indicate 9 low-extinction fields that overlap with MOA.
Magenta boxes show the MOA fields, and the purple regions indicate 
the nominal Roman RGBTDS candidate fields (over-guide).
Sgr A* is located on GB94.
}
\end{center}
\end{figure}

We observe the GB fields when they are visible, from February to October, 
which accounts for approximately 50\% of the total available telescope time over the year.
Figure \ref{fig:fields} shows the PRIME observational fields toward the GB.
The microlensing event rate map in the $I$-band derived from OGLE survey \citep{Mroz19} 
is overlaid as a color map.
The black square grids and associated labels indicate the PRIME observational fields, labeled as GB\#.
The blue squares represent the 35 primary GB fields.
The cyan lines indicate 9 low-extinction fields that overlap with MOA survey fields, 
which are shown in magenta. The purple regions denote the over-guide candidate fields
 for the Roman Space Telescope's RGBTDS \citep{Zasowski2025}.
The Galactic coordinates of all these fields are listed in Table~\ref{tbl:fields}.

\begin{figure}
\begin{center}
\includegraphics[scale=0.45,keepaspectratio]{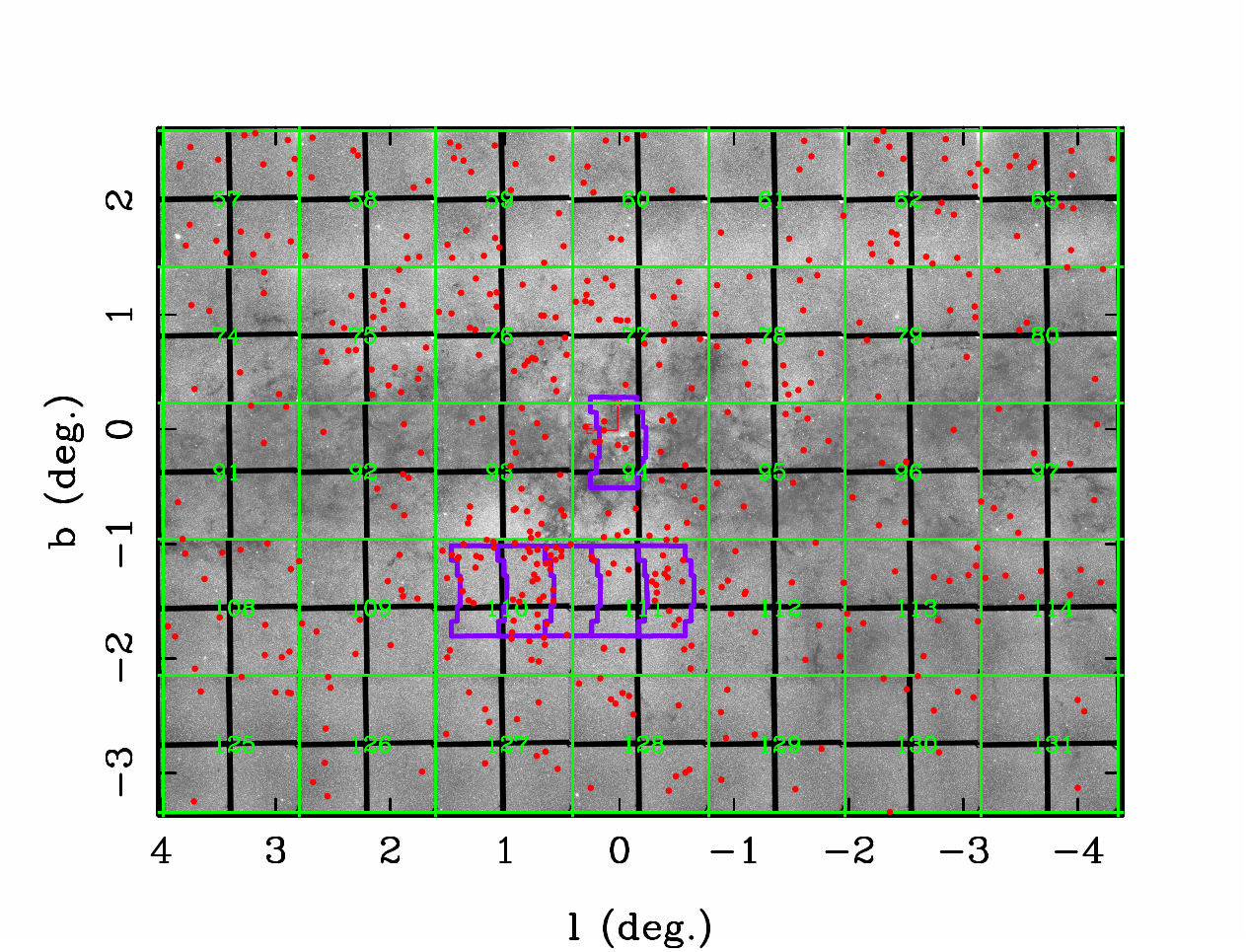}
\caption{
  \label{fig:Mosaic_fields}
Mosaic images of the 35 primary GB fields.
Galactic north is up, and east is to the left.
The field boundaries are outlined in green and labeled with their field numbers (GB\#). 
Sgr A* is located on chip 2 of GB94.
Black lines represent chip gaps.
Red dots show the positions of the 486 microlensing candidates.
Purple regions indicate the nominal Roman RGBTDS candidate fields (over-guide).
%The green lines indicates the observational fields of MOA in optical.
}
\end{center}
\end{figure}

\begin{figure}
\begin{center}
\includegraphics[scale=0.5,keepaspectratio]{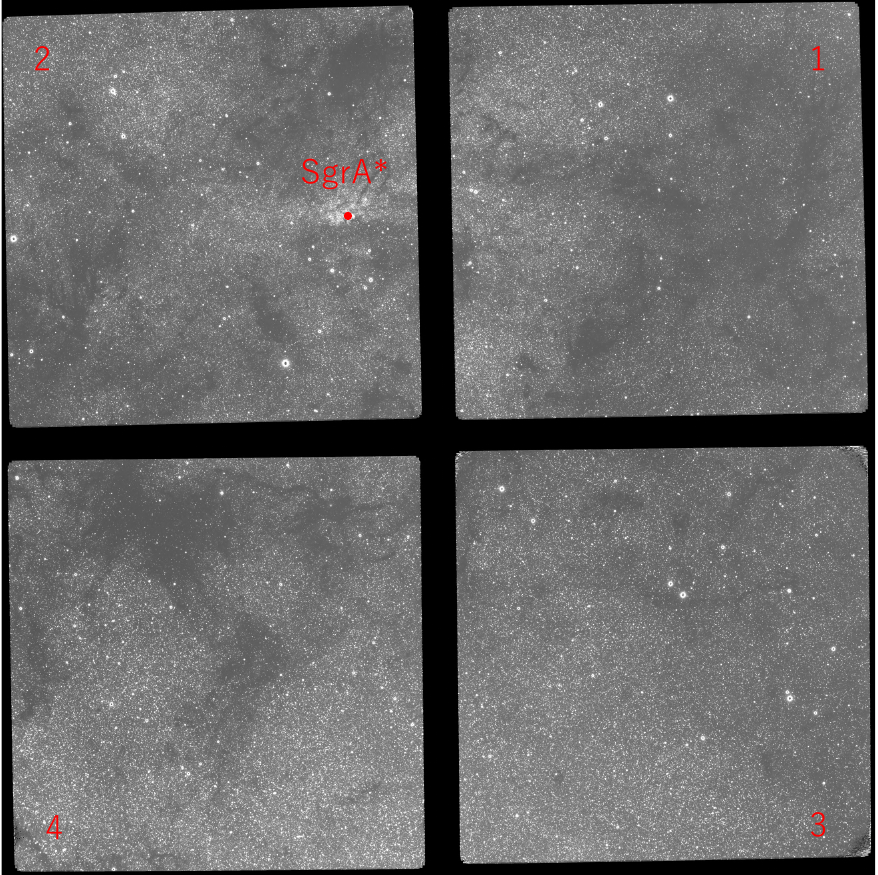}
\caption{
  \label{fig:Image-H4SgrA}
Images of the field GB94 with its 4 chips.
Galactic north is up, and east is to the left.
Chip numbers are shown in red.
Sgr A* is located on chip 2.
}
\end{center}
\end{figure}

\begin{figure}
\begin{center}
\includegraphics[scale=0.5,keepaspectratio]{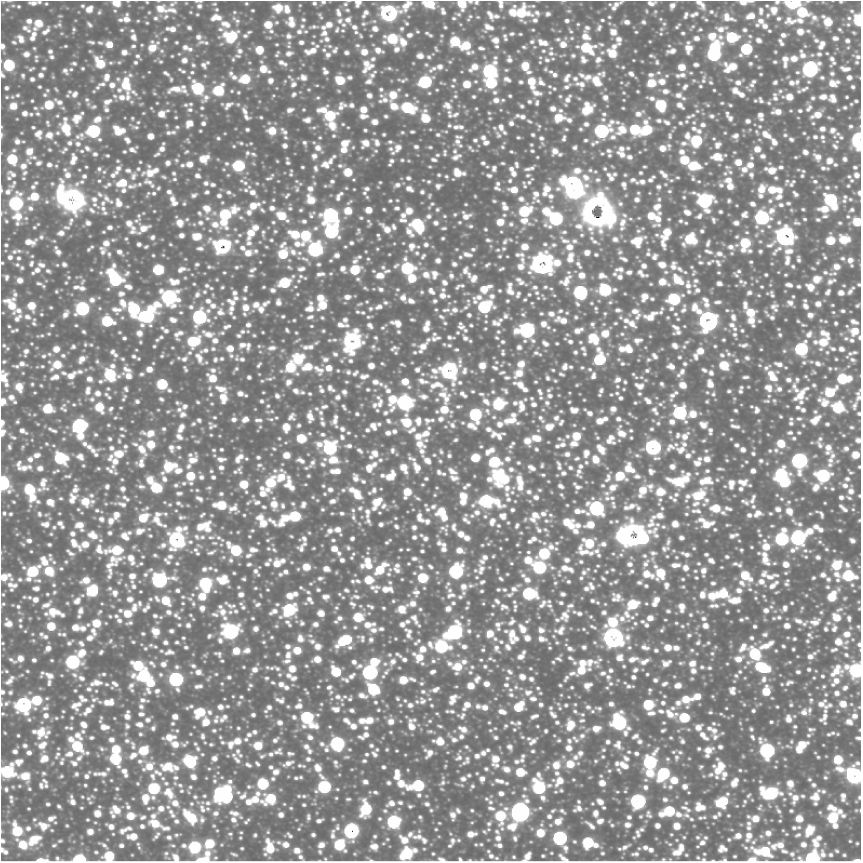}
\caption{
  \label{fig:image_zoom}
The 2000$\times$2000pixel sub-image of GB110,
one of the most stellar dense field, with 8.58 second exposure.
}
\end{center}
\end{figure}

From the start of the 2024 season until June 1, 2025, 
we primarily observed the 35 primary GB fields in the $H$ band.
We take six exposures per visit with a dithering radius of 10 arcseconds.
Each exposure consists of four frames, of which three are effectively usable
since the first frame (frame 0) is used to remove the bias.
The total effective exposure time for six exposures is $2.86 \times 3 \times 6 = 51$ seconds.
Including overhead, each visit takes approximately 140 seconds, 
resulting in about 82 minutes to complete one full cycle of the 35 fields.
%overhead is 175\%
This corresponds to approximately 4-9 visits per field per night.
(Note that prior to August 2024, we used 12 exposures per visit—resulting in an effective exposure time
of 103 seconds—without dithering, corresponding to a cadence of 164 minutes.)
Figure~\ref{fig:Mosaic_fields} shows a mosaic of the observed images of the 35 primary fields, 
along with the positions of 486 detected microlensing candidates compared to the candidate Roman fields.
Figure~\ref{fig:Image-H4SgrA} displays the four-chip image of field GB94, which includes Sgr A* at the Galactic center in chip 2.
Figure~\ref{fig:image_zoom} shows a 2000~$\times$~2000 pixel sub-image of GB110, 
one of the most densely populated stellar fields.

We also observed these fields in the $J$ band with a cadence of approximately three nights 
to obtain the $J-H$ color of the sources, which is useful for characterizing their angular radii \citep{Boyajian2014}.
In addition, we observed nine low-extinction fields in the $H$-band with a three-day cadence, 
overlapping with optical observations by MOA, in order to measure the $H$-band magnitudes of the sources.

Note that, since June 2, 2025, we have modified our observational strategy. See more details in 
the Appendix \ref{sec-append}.

%------------------------Table 4.---------------------------------
\begin{deluxetable}{lrr}
\tabletypesize{\scriptsize}
%\rotate
\tablecaption{Observational fields of PRIME towards the GB in the galactic coordinates.
\label{tbl:fields}
}
\tablewidth{0pt}
\tablehead{
\colhead{ Field } &
\colhead{$l$ ($^\circ$) } &
 \colhead{$b$ ($^\circ$) }
%\colhead{} & 
%\colhead{$(\rm day)$} & 
%\colhead{} & 
%\colhead{} 
}
\startdata
 GB57 &   3.38681 &   2.01403 \\
 GB58 &   2.19681 &   2.01403 \\
 GB59 &   1.00681 &   2.01403 \\
 GB60 &  -0.18319 &   2.01403 \\
 GB61 &  -1.37319 &   2.01403 \\
 GB62 &  -2.56319 &   2.01403 \\
 GB63 &  -3.75319 &   2.01403 \\
 GB74 &   3.38681 &   0.82403 \\
 GB75 &   2.19681 &   0.82403 \\
 GB76 &   1.00681 &   0.82403 \\
 GB77 &  -0.18319 &   0.82403 \\
 GB78 &  -1.37319 &   0.82403 \\
 GB79 &  -2.56319 &   0.82403 \\
 GB80 &  -3.75319 &   0.82403 \\
 GB91 &   3.38681 &  -0.36597 \\
 GB92 &   2.19681 &  -0.36597 \\
 GB93 &   1.00681 &  -0.36597 \\
 GB94 &  -0.18319 &  -0.36597 \\
 GB95 &  -1.37319 &  -0.36597 \\
 GB96 &  -2.56319 &  -0.36597 \\
 GB97 &  -3.75319 &  -0.36597 \\
GB108 &   3.38681 &  -1.55597 \\
GB109 &   2.19681 &  -1.55597 \\
GB110 &   1.00681 &  -1.55597 \\
GB111 &  -0.18319 &  -1.55597 \\
GB112 &  -1.37319 &  -1.55597 \\
GB113 &  -2.56319 &  -1.55597 \\
GB114 &  -3.75319 &  -1.55597 \\
GB125 &   3.38681 &  -2.74597 \\
GB126 &   2.19681 &  -2.74597 \\
GB127 &   1.00681 &  -2.74597 \\
GB128 &  -0.18319 &  -2.74597 \\
GB129 &  -1.37319 &  -2.74597 \\
GB130 &  -2.56319 &  -2.74597 \\
GB131 &  -3.75319 &  -2.74597 \\
\hline
Low extinction fields\\
\hline
GB124 &   4.57681 &  -2.74597 \\
GB141 &   4.57681 &  -3.93597 \\
GB142 &   3.38681 &  -3.93597 \\
GB143 &   2.19681 &  -3.93597 \\
GB144 &   1.00681 &  -3.93597 \\
GB145 &  -0.18319 &  -3.93597 \\
GB146 &  -1.37319 &  -3.93597 \\
GB147 &  -2.56319 &  -3.93597 \\
GB148 &  -3.75319 &  -3.93597 \\
 \enddata
 %\tablenotetext{a}{Likely Brown dwarf lens.}
%\table comments{
%}
\end{deluxetable}
%----------------------------------------------------------------

%----------------------------- FIG. 5 -------------------------------------
%--------------------------------------------------------------------------

%----------------------------- FIG. 6-------------------------------------
%--------------------------------------------------------------------------

%----------------------------- FIG. 7-------------------------------------
%--------------------------------------------------------------------------

%----------------------------- FIG. 8-------------------------------------
%--------------------------------------------------------------------------

%==================================

\section{Data reduction}
\label{sec:Data_reduction}

The raw frame image are UINT16 and has a size of 34.7 MB ( 4230$\times$4096 pix $\times$ 2 Bytes) per chip, 
and 23-28 MB after compression.
Here 4230 columns include 4096 columns of science pixels, 
6 columns of ACADIA telemetry data at the start of the array and
128 columns of reference voltage data at the end of the array.
This corresponds to approximately  %23\UTF{2013}28 GB per 1,000 frames, and 
about 120 TB per year.
The acquired raw images are compressed and stored on-site, 
then transferred to servers in Japan via the internet and/or external hard drives.
Image reduction follows standard procedures for NIR detectors, 
with our own implementation for non-linearity correction (Hamada et al., in preparation), as described below:
(1) Super-bias subtraction: %A super-bias is created at the same detector temperature as follows.
We acquired 10 SUTR exposures at each of eight different thermal background levels, 
resulting in a total of 80 exposures at the same detector temperature.
The measured fluxes range from 100 to 500 ADU with 17-second exposures, 
corresponding to 6 frames per exposure. For each exposure and pixel, 
a linear fit was performed using frames 1 to 5, excluding the first frame (frame 0).
We determined the intersection point of the extrapolated fitted lines—this point is 
where the RMS is minimized for each exposure and pixel.
The super-bias is defined as the average ADU value at these intersection points.
This super-bias is then subtracted from the raw science image frames.
(2) Reference pixel correction: 
The detector is read out through 32 channels, each with a size of 128$\times$4096 pixels and similar bias levels. Reference pixels form a 4-pixel-wide border along the top, bottom, left, and right edges of the 4088$\times$4088 active pixel area. To remove column-dependent bias, the median of the 128$\times$4 reference pixels located at the top and bottom of each output channel should be subtracted from each column. Here we don't use the reference pixels at the bottom for the first frame, frame 0, as these are unstable. 
Here, we use only the top reference pixels for frame 0, while for frames 1, 2, and 3, we use both the top and bottom reference pixels.
(3) Non-linearity correction: Infrared detectors are known to exhibit non-linearity.
Linearity curves, obtained at similar detector temperatures, are used to correct each science frame.
This procedure is similar to that used for the H2RG detector in JWST \citep{Canipe2017}, 
but is based on our own implementation (Hamada et al., in preparation).
(4) Dark subtraction: Dark frames are created from the slopes of 10 frames (corresponding to 28.6 s exposure) taken with the blocked position. These dark frames are then subtracted from each science frame.
(5) Up-the-ramp fitting: Final science images are produced using up-the-ramp fitting.
We refer to these corrected images as “slope images.” which is UNIT 32 and has a size of 67.1 MB per chip, 
(6) Flat-fielding: Twilight flat images are taken before and/or after each night’s observations.
Pixel response variations are corrected using these flat-field frames.
(7) Stacking: The exposures obtained in a single visit are combined using median stacking to produce deep images. This process effectively removes most cosmic-ray hits and bad pixels. The stacked images are used for event detection, while the original unstacked images are also analyzed to preserve higher time resolution.

\begin{figure}
\begin{center}
\includegraphics[scale=0.35,keepaspectratio]{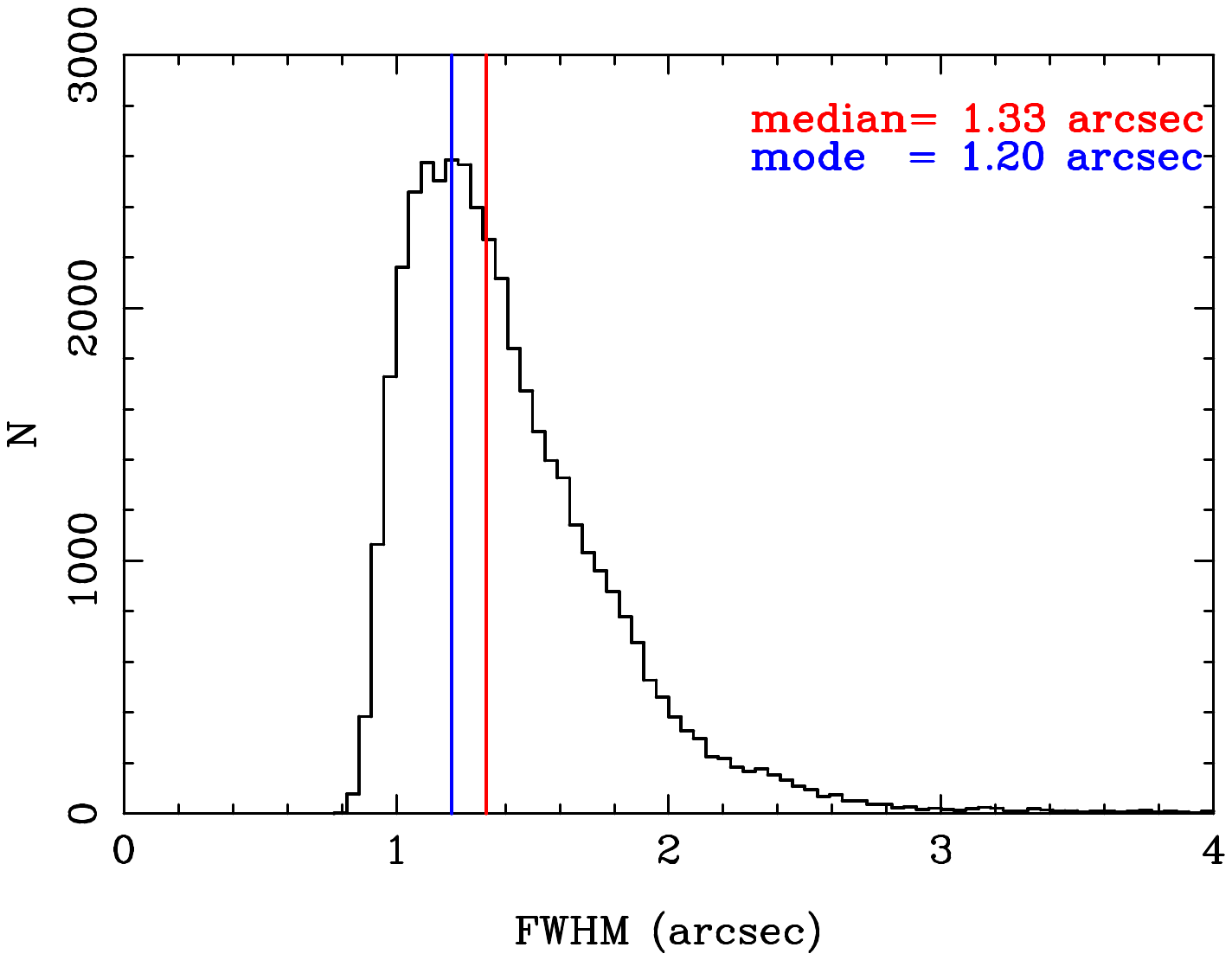}
\caption{
  \label{fig:Seeing}
The distribution of the observed PSF FWHM in the $H$-band images over approximately 14 months.
The median is $1.33''$ (red line), and the mode is $1.20'' $ (blue line).
 }
\end{center}
\end{figure}

We show the distribution of PSF FWHMs in the observed $H$-band images over approximately 14 months in Figure~\ref{fig:Seeing}.
The median FWHM is $1.33$ arcsec, and the mode is $1.20$ arcsec.
This can be explained by the expected natural seeing of $\sim$1.1 arcsec in the $H$-band, estimated from 
Differential Image Motion Monitor (DIMM) measurements of 1.4 arcsec in the optical \citep{Catala2013}, a 50\% EED of 0.45 arcsec (for the as-built optics), 
and additional air turbulence inside the dome, although the latter has not been quantified.

% FWHM in H-band(1.65um) is 1.4 * (1.65um/0.5um)^(-1/5) = 1.1 arcsec.

% In 2D gaussian,  Integral within HWHM is 50% 
% 80% EER= 7.0um (as designed) diameter= 14um 
% 50% EER= 3.0um (as designed) diameter= 14um 
% 80% EER= 8.0um (as biuld)
% 50% EER= 4.5um (as build), Diameter= 9um. --> 9um/10um/pix = 0.9pix --> 0.9pix *0.5arcsec/pix = 0.45arcsec
%1.1^2 +0.45^2 = 1.19 arcsec

%----------------------------- FIG. 9 -------------------------------------
%--------------------------------------------------------------------------

We process the slope images using Difference Image Analysis (DIA) \citep{bon01}.
The reference images are constructed by stacking 12 exposures, 
corresponding to a total exposure time of 103 seconds per field.
DIA can be performed within 5 minutes of acquiring the images.

\begin{figure}
\begin{center}
\includegraphics[scale=0.35,keepaspectratio]{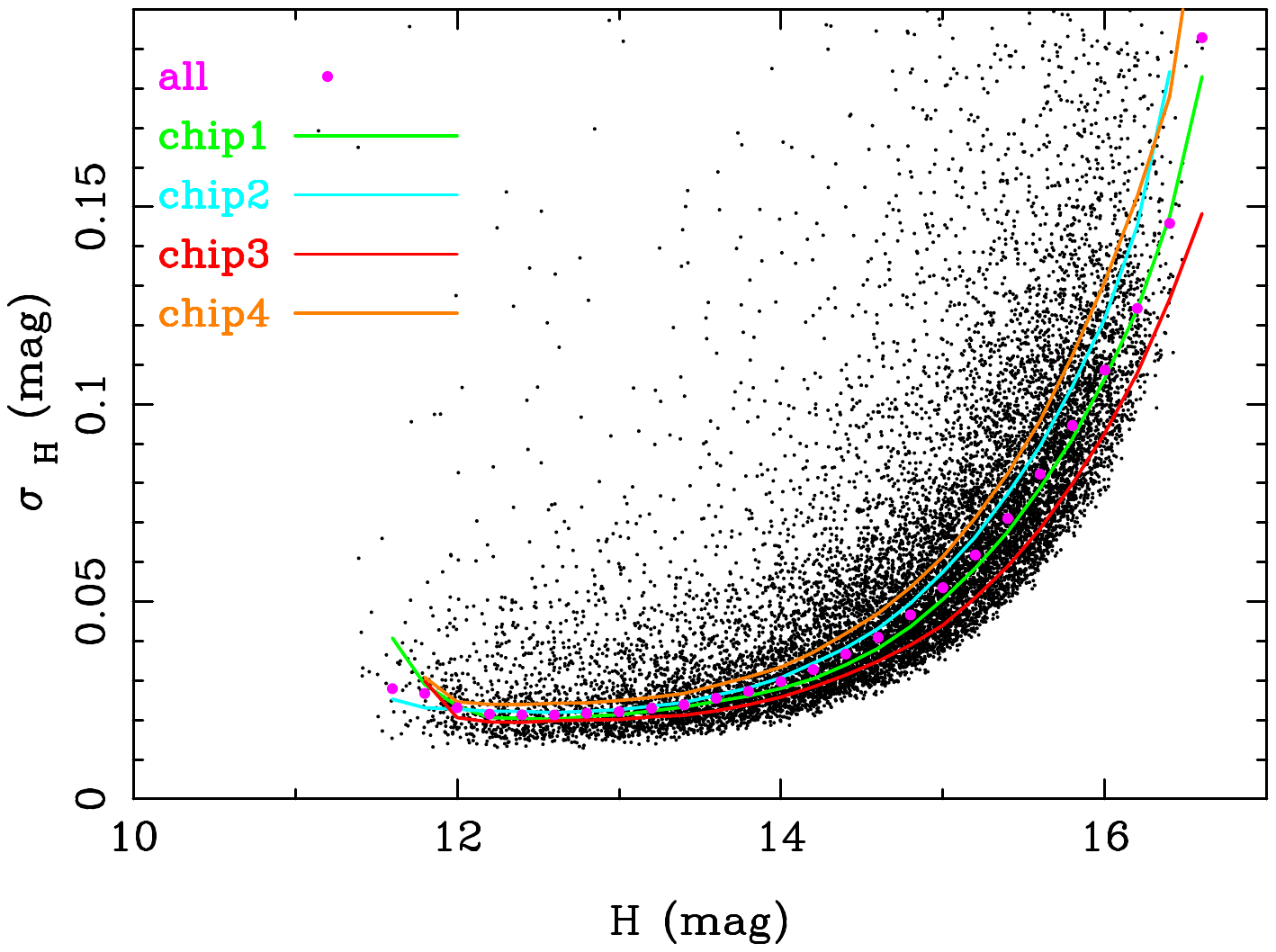}
\caption{
  \label{fig:sigma_H}
Photometric precision in the GB fields as a function of $H$-band magnitude.
The black points show the standard deviation, $\sigma_{H}$, of light curves for 
a sub-sample of 69,000 non-variable stars in the GB94 field.
Outliers with larger $\sigma_{H}$ values are likely due to remaining variable objects.
The green, cyan, red, and orange lines indicate the median values in magnitude bins 
for chips 1, 2, 3, and 4, respectively, while the magenta filled circles represent the median over all chips.
 }
\end{center}
\end{figure}

 The light curves are generated using PSF photometry on the difference images.
The PSFs are estimated from the reference images and then convolved 
with the DIA kernel for each image to match the FWHM.
The photometric precision is assessed by calculating the standard deviation ($\sigma_{H}$) 
in the $H$-band light curves of about 69,000 non-variable stars that are resolved 
in the reference images within the GB94 field.
In Figure~\ref{fig:sigma_H}, black does show $\sigma_{H}$ for a sub-sample of 
these stars as a function of $H$-band magnitude.
The green, cyan, red, and orange lines indicate the median values 
of all sample as a function of $H$-band magnitude for chips 1, 2, 3, and 4, respectively.
The magenta filled circles represent the median over all four chips.
Table~\ref{tbl:sigma_H} summarizes the median and mode of $\sigma_{H}$ 
as a function of $H$-band magnitude for each chip and for the combined sample.
Chip 3 exhibits the best precision, while Chip 4 performs the worst, reflecting their detector characteristics.
The lowest median and mode values are about 0.019-0.024 mag and 0.016-0.020 mag at $H=12$-13 mag, respectively.
For brighter magnitudes, $\sigma_{H}$ increases slightly, likely due to systematics and/or intrinsic stellar variability.
For fainter magnitudes, $\sigma_{H}$ increases rapidly due to the high stellar density in the GB fields.

%----------------------------- FIG. 10 -------------------------------------
%--------------------------------------------------------------------------

%------------------------Table 5.---------------------------------
\begin{deluxetable*}{rrrrrrrrrrr}
\tabletypesize{\scriptsize}
%\rotate
\tablecaption{Photometric precision in GB fields as a function of $H$-band magnitude.
\label{tbl:sigma_H}
}
\tablewidth{0pt}
\tablehead{
\colhead{ $H$ } &
\multicolumn{2}{c}{Chip1} & 
\multicolumn{2}{c}{Chip2} &
\multicolumn{2}{c}{Chip3} &
\multicolumn{2}{c}{Chip4} &
\multicolumn{2}{c}{All} \\
\colhead{}  &
\colhead{med}  &
\colhead{mod}  &
\colhead{med}  &
\colhead{mod}  &
\colhead{med}  &
\colhead{mod}  &
\colhead{med}  &
\colhead{mod}  &
\colhead{med}  &
\colhead{mod}  \\
\colhead{(mag)}  &
\colhead{(mag)}  &
\colhead{(mag)}  &
\colhead{(mag)}  &
\colhead{(mag)}  &
\colhead{(mag)}  &
\colhead{(mag)}  &
\colhead{(mag)}  &
\colhead{(mag)}  &
\colhead{(mag)}  &
\colhead{(mag)}  
%\colhead{} & 
%\colhead{$(\rm day)$} & 
%\colhead{} & 
%\colhead{} 
}
\startdata
%\multicolumn{3}{c}{Cadence: 21.6min (4obs./cycle)} \\
\hline
11.6 &  0.041 &  0.038 &  0.025 &  0.020 &  0.000 &  0.000 &  0.000 &  0.000 &  0.028 &  0.020\\
11.8 &  0.029 &  0.028 &  0.023 &  0.020 &  0.030 &  0.026 &  0.031 &  0.026 &  0.027 &  0.020\\
12.0 &  0.024 &  0.022 &  0.023 &  0.020 &  0.021 &  0.018 &  0.025 &  0.022 &  0.023 &  0.020\\
12.2 &  0.021 &  0.018 &  0.022 &  0.018 &  0.020 &  0.018 &  0.024 &  0.020 &  0.021 &  0.018\\
12.4 &  0.020 &  0.018 &  0.022 &  0.018 &  0.019 &  0.018 &  0.024 &  0.020 &  0.021 &  0.018\\
12.6 &  0.020 &  0.018 &  0.022 &  0.018 &  0.020 &  0.016 &  0.024 &  0.020 &  0.021 &  0.018\\
12.8 &  0.021 &  0.018 &  0.022 &  0.018 &  0.020 &  0.018 &  0.024 &  0.020 &  0.022 &  0.018\\
13.0 &  0.021 &  0.020 &  0.023 &  0.018 &  0.020 &  0.018 &  0.025 &  0.022 &  0.022 &  0.020\\
13.2 &  0.022 &  0.020 &  0.023 &  0.020 &  0.021 &  0.018 &  0.026 &  0.022 &  0.023 &  0.020\\
13.4 &  0.023 &  0.020 &  0.024 &  0.022 &  0.021 &  0.018 &  0.027 &  0.024 &  0.024 &  0.020\\
13.6 &  0.025 &  0.022 &  0.026 &  0.022 &  0.022 &  0.020 &  0.029 &  0.026 &  0.026 &  0.022\\
13.8 &  0.026 &  0.024 &  0.028 &  0.024 &  0.024 &  0.022 &  0.031 &  0.026 &  0.027 &  0.024\\
14.0 &  0.028 &  0.026 &  0.031 &  0.026 &  0.026 &  0.024 &  0.033 &  0.028 &  0.030 &  0.026\\
14.2 &  0.030 &  0.028 &  0.034 &  0.030 &  0.028 &  0.026 &  0.037 &  0.032 &  0.033 &  0.028\\
14.4 &  0.034 &  0.032 &  0.038 &  0.034 &  0.031 &  0.028 &  0.042 &  0.034 &  0.037 &  0.032\\
14.6 &  0.038 &  0.036 &  0.043 &  0.040 &  0.035 &  0.032 &  0.047 &  0.040 &  0.041 &  0.036\\
14.8 &  0.044 &  0.042 &  0.050 &  0.046 &  0.039 &  0.036 &  0.054 &  0.044 &  0.047 &  0.044\\
15.0 &  0.051 &  0.048 &  0.057 &  0.054 &  0.044 &  0.040 &  0.061 &  0.052 &  0.054 &  0.050\\
15.2 &  0.058 &  0.056 &  0.066 &  0.062 &  0.051 &  0.048 &  0.071 &  0.062 &  0.062 &  0.058\\
15.4 &  0.068 &  0.062 &  0.077 &  0.074 &  0.059 &  0.056 &  0.082 &  0.072 &  0.071 &  0.068\\
15.6 &  0.079 &  0.074 &  0.089 &  0.084 &  0.068 &  0.066 &  0.096 &  0.082 &  0.082 &  0.080\\
15.8 &  0.091 &  0.088 &  0.105 &  0.100 &  0.080 &  0.078 &  0.112 &  0.098 &  0.095 &  0.092\\
16.0 &  0.107 &  0.104 &  0.122 &  0.116 &  0.093 &  0.092 &  0.131 &  0.118 &  0.109 &  0.102\\
16.2 &  0.124 &  0.120 &  0.145 &  0.132 &  0.108 &  0.104 &  0.153 &  0.132 &  0.124 &  0.116\\
16.4 &  0.148 &  0.136 &  0.184 &  0.160 &  0.127 &  0.124 &  0.178 &  0.170 &  0.146 &  0.124\\
16.6 &  0.183 &  0.182 &  0.000 &  0.000 &  0.148 &  0.142 &  0.232 &  0.186 &  0.193 &  0.142\\
 \enddata
 %\tablenotetext{a}{Likely Brown dwarf lens.}
%\table comments{
%}
\end{deluxetable*}
%----------------------------------------------------------------

New events are identified
 following a similar procedure used in the MOA survey \citep{bon01}.
On the subtracted images, possible variable star objects are detected using 
a custom implementation of the IRAF task DAOFIND \citep{daophot} 
that is adapted for difference images.
Light curves are then generated for the detected objects.
Simple preliminary cuts are applied to select events showing instantaneous magnification 
relative to the baseline and to remove a large fraction of artifacts, while known variable stars are masked.
Among the remaining objects, microlensing candidates are selected by visual inspection, 
and the list of variable stars is continuously updated for masking purposes.
Finally, alerts of the microlensing candidates are publicly posted on the web.
All artifacts are rejected and obvious variable stars, while not alerted, are tagged for further studies.
Some irregular variables, transient variables, such as cataclysmic variables (CVs) or stellar flares, 
also feature on the alert list because they are difficult to distinguish at the onset of the brightening.
Note that the preliminary cuts are empirically refined to improve efficiency.
Therefore, an offline analysis with fixed thresholds should be applied for statistical studies.

\begin{figure}
\begin{center}
\includegraphics[scale=0.37,keepaspectratio]{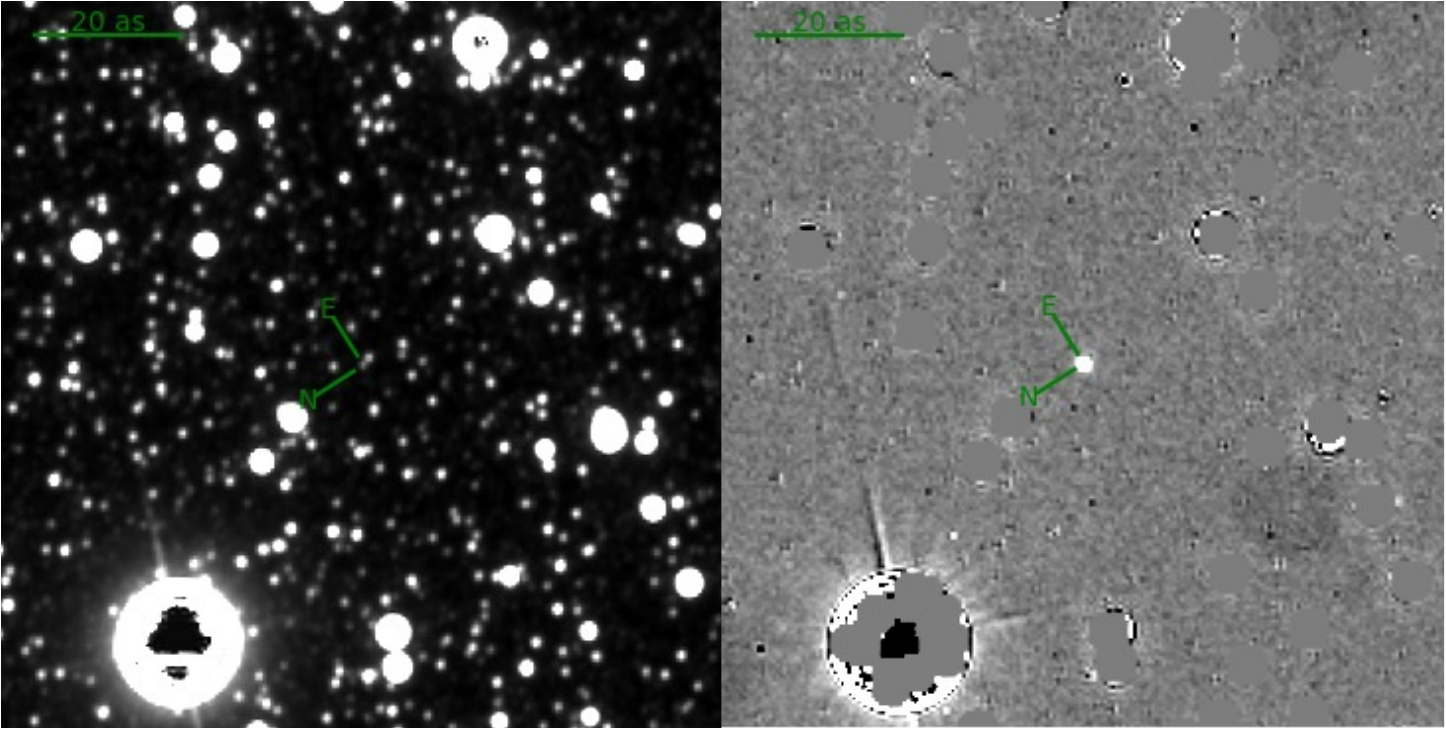}
\includegraphics[scale=0.37,keepaspectratio]{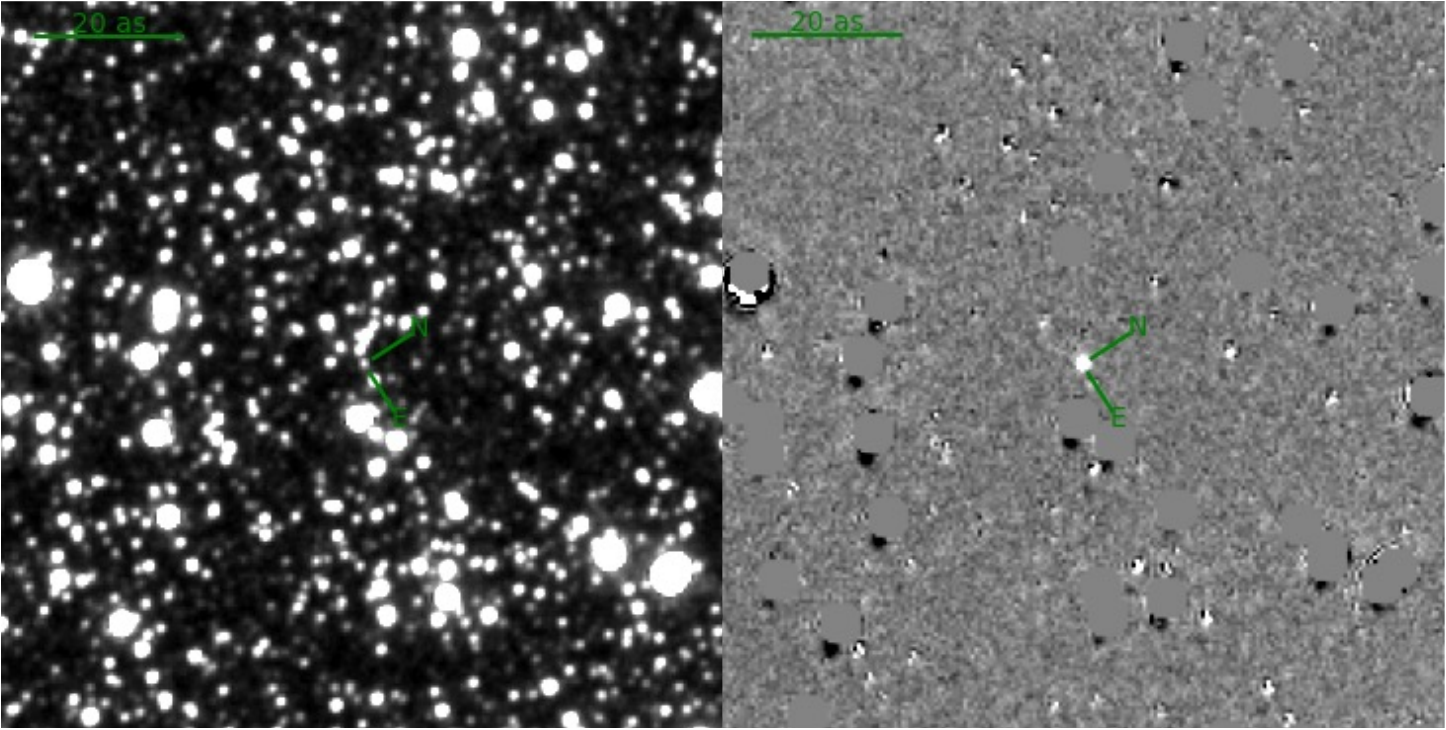}
\caption{
  \label{fig:image_diff}
The slope images of the microlensing events PRIME-2024-BLG-124 (top panels) and 170 (bottom panels)
before (left) and after (right) DIA processing with  50$\times$50 arcsecs (100$\times$100 pixels).
The targets are marked with a green cross.
}
\end{center}
\end{figure}

\begin{figure}[] 
\begin{center}
\includegraphics[scale=0.33,keepaspectratio]{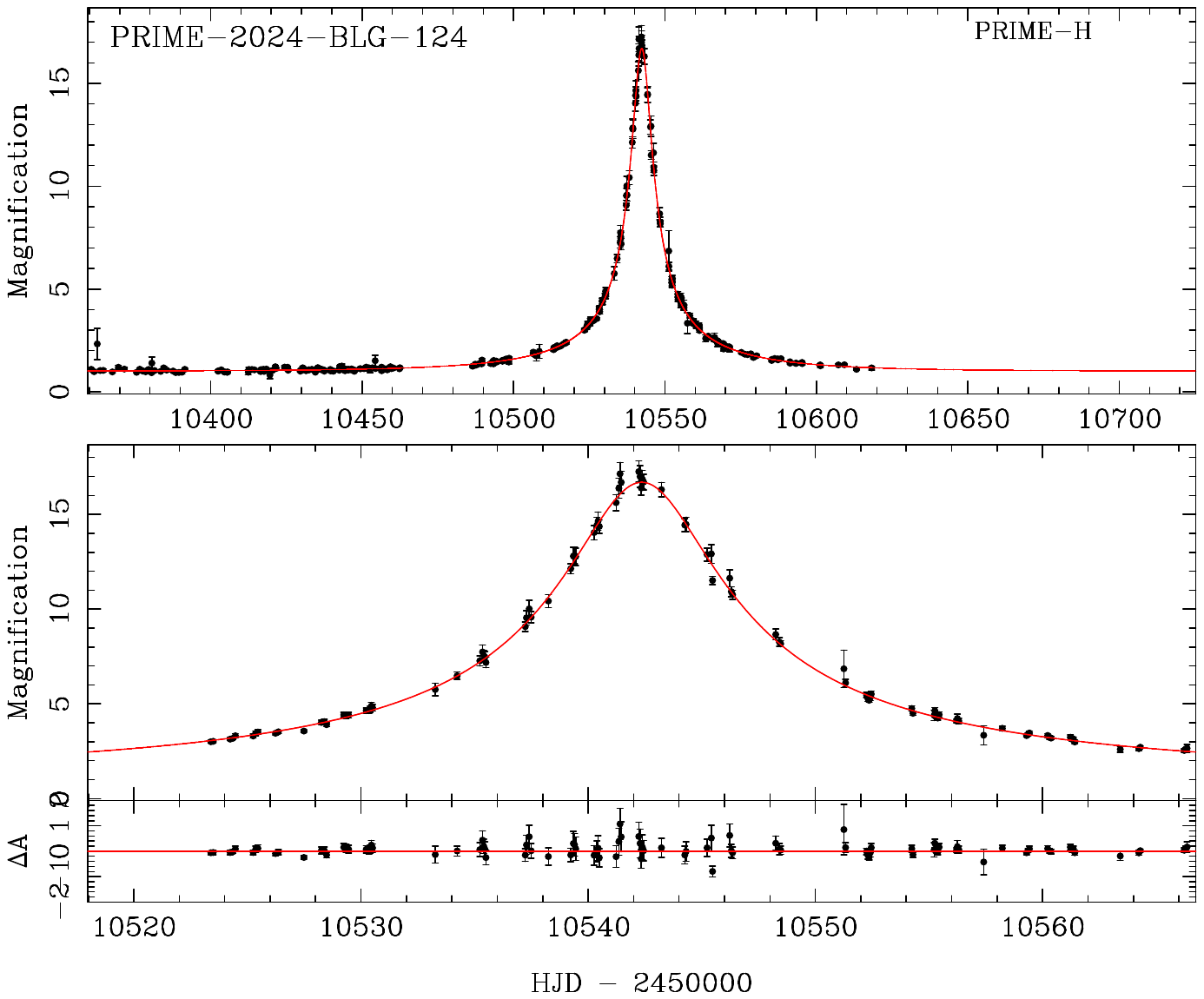}
\includegraphics[scale=0.33,keepaspectratio]{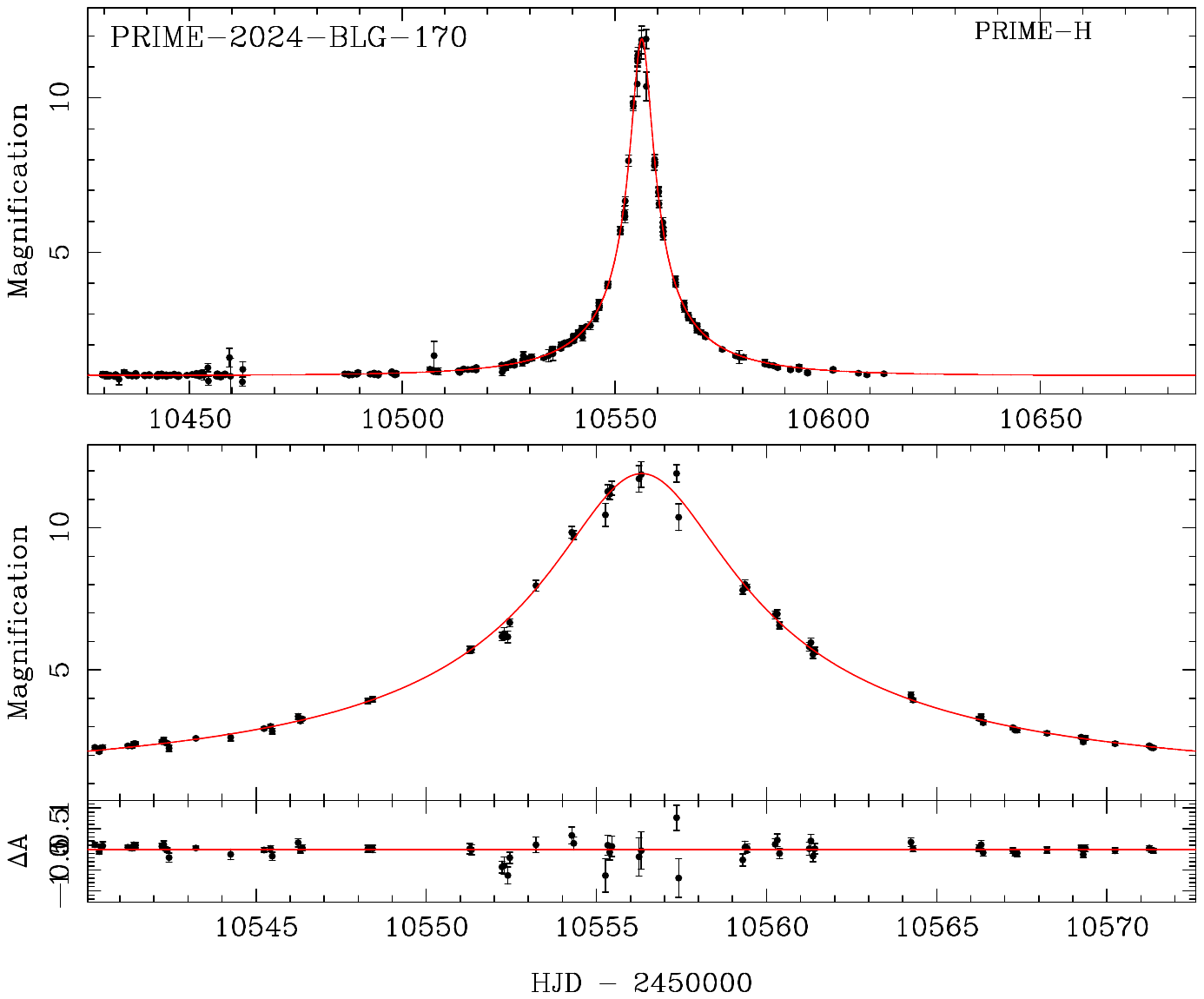}
\caption{
  \label{fig:lc}
The light curve of PRIME-2024-BLG-124 and 170, detected in field GB75 and GB78, respectively.
The vertical axis shows the differential flux ($\Delta$flux) in ADU.
The red lines indicate the best-fit models, 
with event timescales of $t_{\rm E} = 56.5 \pm 2.0$ days and $32.2 \pm 0.6$ days and
and impact parameters of $u_0 = 0.060 \pm 0.0029$ and $0.084 \pm 0.002$ 
and $H$-band source magnitudes of $H_s=12.90\pm0.20$ mag and $H_s=14.90\pm0.20$ mag, respectively.
%$t_{\rm E} = 14.9$ days, $u_0 = 0.14$, and $H_{\rm s} = 12.9$ mag, respectively.
}
\end{center}
\end{figure}

Figure \ref{fig:image_diff} shows the slope images before and after DIA for the microlensing events 
PRIME-2024-BLG-124 and PRIME-2024-BLG-170, 
which were detected in the field GB75 and GB78, respectively.
Figure \ref{fig:lc} displays the light curve of these events.
%https://moaprime.massey.ac.nz/prime2024

%----------------------------- FIG. 11-------------------------------------
%--------------------------------------------------------------------------

%----------------------------- FIG. 12-------------------------------------
%--------------------------------------------------------------------------

We have preliminarily identified 284 microlensing candidates 
during the 2024 survey season\footnote{\tt https://moaprime.massey.ac.nz/prime2024} and 
202 candidates up to June 1, 2025\footnote{\tt https://moaprime.massey.ac.nz/alerts/index/prime/2025}
before changing the strategy (See the Appendix \ref{sec-append}). 
The list of the microlensing candidates are shown in Table \ref{tbl:events} and full list is available electrically.
The positions of these 486 candidates are shown in Figure~\ref{fig:Mosaic_fields} for comparison with 
the candidate Roman fields. One of the purposes of the PRIME bulge survey is to provide the event distribution
as input to help define the Roman survey fields. Although the current statistics are still insufficient, 
our results indicate that the event rate is relatively higher in the current candidate fields of Roman.

\begin{figure}
\begin{center}
\includegraphics[scale=0.33,keepaspectratio]{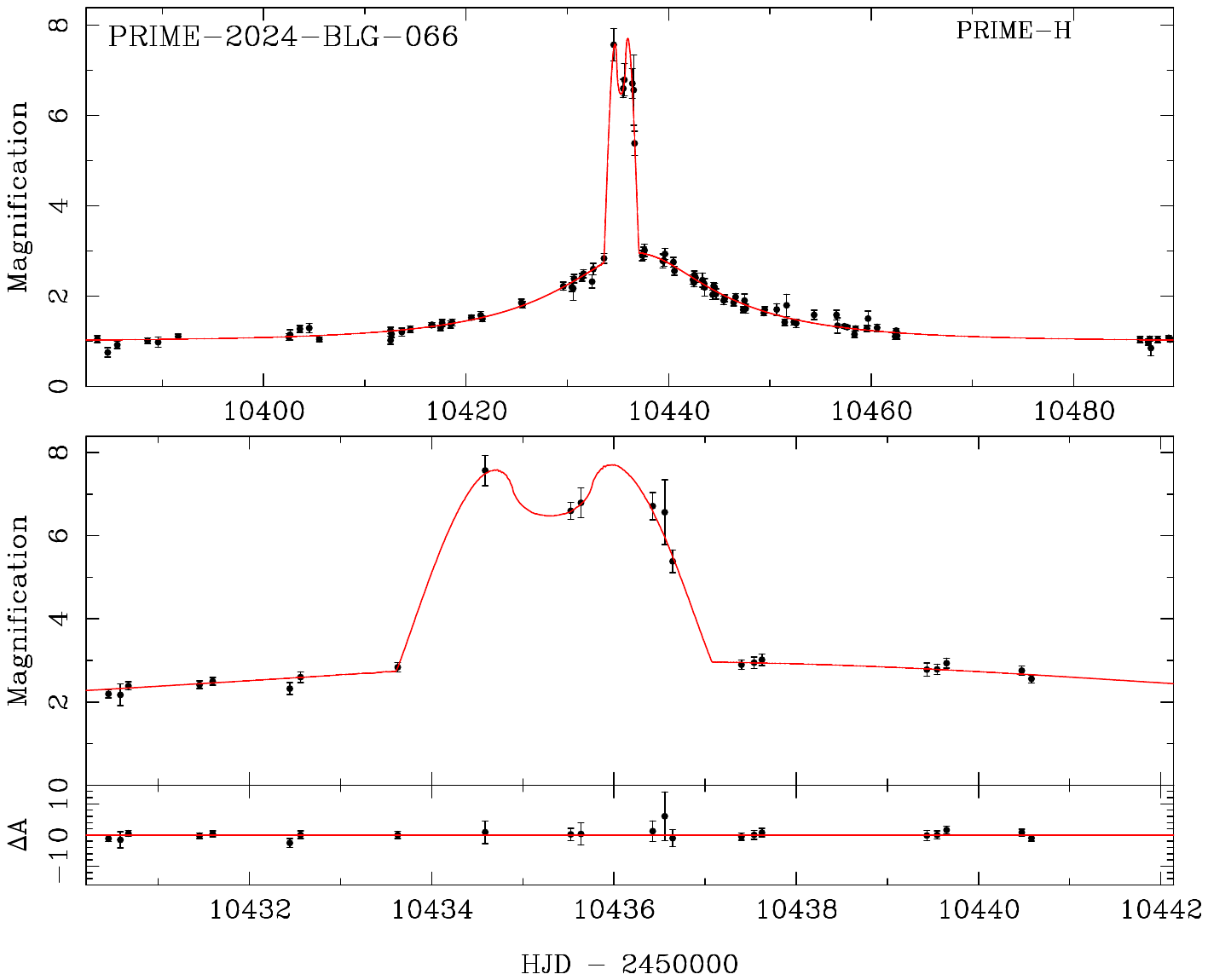}
\caption{
  \label{fig:lc_binary}
The light curve of a binary event PRIME-2024-BLG-066, detected in field GB92.
The vertical axis shows the differential flux ($\Delta$flux) in ADU.
The best fit parameters are the mass ratio $q=0.139\pm0.019$, 
separation $s=1.531\pm0.076$, 
event timescale $t_{\rm E} = 21.8\pm2.9$ days. 
% and impact parameter $u_0 = 0.228 ± 0.040$.
%and H-band source magnitude are  and 
%$H_{\rm s} = 15.3$ mag, 
%respectively.
}
\end{center}
\end{figure}

Among these candidates, 41 events exhibit deviations from the point-source point-lens (PSPL) model, 
referred to as "anomalies." Figure~\ref{fig:lc_binary} shows an example light curve of 
the binary microlensing event PRIME-2024-BLG-066. 
The best-fit parameters are a mass ratio of $q = 0.139 \pm 0.019$, 
a separation of $s = 1.531 \pm 0.076$, and an event timescale of $t_{\rm E} = 21.8 \pm 2.9$ days. 
A full analysis of these anomalous events is beyond the scope of this paper.

% https://moaprime.massey.ac.nz/alerts

%----------------------------- FIG. 10-------------------------------------
%\begin{figure}
%\begin{center}
%\includegraphics[scale=0.5,keepaspectratio]{Fig/PRIMEH_PB24124.dat.flux.PSPL5c.pdf}
%\includegraphics[scale=0.5,keepaspectratio]{Fig/PRIMEH_PB24163.dat.flux.PSPL5c.pdf}
%\includegraphics[scale=0.5,keepaspectratio]{Fig/PRIMEH_PB24170.dat.flux.PSPL5c.pdf}
%\caption{
%  \label{fig:lc}
%}
%\end{center}
%\end{figure}
%--------------------------------------------------------------------------
%------------------------Table 6.---------------------------------
\begin{deluxetable*}{lrrrrrrrr}
\tabletypesize{\scriptsize}
%\rotate
\tablecaption{Microlensing candidates from 2024 season.
\label{tbl:events}
}
\tablewidth{0pt}
\tablehead{
\colhead{ ID } &
\colhead{ R.A. } &
\colhead{Dec.} &
\colhead{ $l$ } &
\colhead{$b$} &
\colhead{$t_{\rm 0}$} &
\colhead{$t_{\rm E}$} &
\colhead{$u_{\rm 0}$} &
\colhead{$H_{\rm 0}$} \\
\colhead{ } &
\colhead{(2000)} &
\colhead{(2000)} &
\colhead{(deg.)} &
\colhead{(deg.)} &
\colhead{(HJD-2450000)} &
\colhead{(day)} &
\colhead{} &
\colhead{(mag)} 
%\colhead{} & 
%\colhead{$(\rm day)$} & 
%\colhead{} & 
%\colhead{} 
}
\startdata
%PRIME-2024-BLG-001 & +17:58:38.0960 & -27:33:35.9680 & 553.7730 &    54.92 &   0.1053 &    16.73\\
%PRIME-2024-BLG-002 & +17:54:22.4542 & -27:51:04.9599 & 559.3619 &    32.23 &   0.0014 &    18.35\\
%PRIME-2024-BLG-003 & +17:57:20.7517 & -26:37:48.7847 & 354.3834 &    28.09 &   0.7840 &    14.12\\
%PRIME-2024-BLG-004 & +17:38:04.0226 & -27:32:01.3268 & 351.3146 &    39.17 &   0.2165 &    13.80\\
%PRIME-2024-BLG-005 & +17:52:13.8296 & -28:31:55.2943 & 359.8949 &    19.64 &   0.1389 &    16.73\\
PRIME-2024-BLG-001 & +17:58:38.0960 & -27:33:35.9680 &  2.64501 & -1.76469  & 10553.7730 &    54.92 &   0.1053 &    16.73\\
PRIME-2024-BLG-002 & +17:54:22.4542 & -27:51:04.9599 &  1.91929 & -1.09544  & 10559.3619 &    32.23 &   0.0014 &    18.35\\
PRIME-2024-BLG-003 & +17:57:20.7517 & -26:37:48.7847 &  3.30717 & -1.05153  & 10354.3834 &    28.09 &   0.7840 &    14.12\\
PRIME-2024-BLG-004 & +17:38:04.0226 & -27:32:01.3268 &  0.31354 &  2.15309  & 10351.3146 &    39.17 &   0.2165 &    13.80\\
PRIME-2024-BLG-005 & +17:52:13.8296 & -28:31:55.2943 &  1.09294 & -1.03386  & 10359.8949 &    19.64 &   0.1389 &    16.73\\
 \enddata
 %\tablenotetext{a}{Likely Brown dwarf lens.}
\tablecomments{The full list is available electrically.}
%}
%\caption{基本的な表}
\end{deluxetable*}
%----------------------------------------------------------------

%----------------------------- FIG. 13-------------------------------------
%--------------------------------------------------------------------------

\section{ALL-SKY GRID FIELDS}
\label{sec:ALL-SKY}
%The PRIME telescope also conducts Target-of-Opportunity (ToO) observations for various transient events, 
%such as $\gamma$-ray bursts, gravitational wave events, tidal disruption events (TDEs), 
%and others.
%In 2024, we carried out TBC ToO observations (reference), resulting in the detection of TBC counterparts.

\begin{figure}
\begin{center}
\includegraphics[scale=0.45,keepaspectratio]{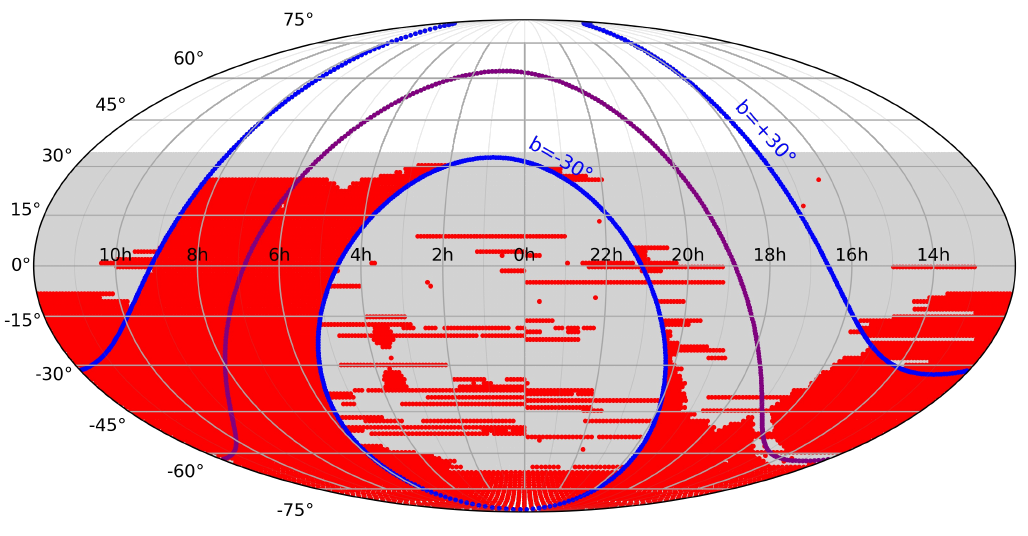}
\caption{
  \label{fig:AllSkyGrid}
PRIME All-Sky Grid fields in Equatorial Coordinates.
The purple and blue lines indicate the Galactic equator and $b = \pm 30^\circ$, respectively.
Red dots show the fields observed as of May 7, 2025, covering 40.8\% 
of the sky accessible to PRIME (shaded in gray).
}
\end{center}
\end{figure}

When the GB fields are not observable 
%and there are no active ToO targets,
we observe all-sky grid fields, which can later serve as reference images for identifying transient events.
As of May 7, 2025, we have observed approximately 13,000 deg$^2$ in the $J$-band (indicated by red dots),
which corresponds to about 41\% of the sky accessible to PRIME shown in gray in Figure~\ref{fig:AllSkyGrid}.

%----------------------------- FIG. 14-------------------------------------
%--------------------------------------------------------------------------

%==================================
\section{Discussion and conclusions}
\label{sec:discussionAndSummary}
 
PRIME is the first dedicated wide-field NIR microlensing survey telescope, located at the Sutherland Observatory in South Africa.
It features one of the widest fields of view among NIR telescopes of its class.
The telescope and camera were installed in 2022, and observations began in mid-2023.
We have preliminarily analyzed the images from the start of the 2024 season until June 1, 2025, 
and identified  486 microlensing candidates, 
along with more than a thousand variable stars (Matsunaga et al. in preparations).
Although the current statistics are still insufficient, the spatial distributions of  these 486 candidates 
indicate that the event rate is relatively higher in the current candidate fields of Roman.

We are currently working on improving the data analysis pipeline by fine-tuning it for the characteristics of the PRIME images.
We continue to issue real-time alerts for microlensing candidates, 
which will encourage follow-up observations by other telescopes for exoplanet detection.
One of the primary scientific goals of PRIME is to study the frequency of exoplanets in the central GB 
and compare it with that in the outer bulge.

Spectroscopic follow-up of highly magnified source stars in the GB is highly valuable 
for studying the chemical evolution of the bulge \citep{Bensby2021}.
Events discovered by PRIME provide opportunities to probe the central regions of the bulge, 
which have been inaccessible to optical surveys.

The Roman GBTDS will carry out a high-cadence NIR survey 
toward the GB as its exoplanet microlensing program.
Roman can observe the GB during the spring and autumn seasons, for up to 72 days per season.
The Roman Observations Time Allocation Committee (ROTAC) recently recommended 
an over-guide allocation for GBTDS, consisting of 12.1-minute cadence observations of six fields 
across six seasons of 70.5 days each, resulting in a total observing time of 438 days 
distributed over the 5-year prime mission \citep{Zasowski2025}.
The remaining four seasons will be covered with a lower cadence of 3-5 days.
After the launch of Roman, PRIME will conduct concurrent observations to measure 
space-based microlensing parallax, enabling the determination of the mass and 
distance of lens systems from Earth.
PRIME will also be able to monitor bright stars during the seasons not covered by Roman, 
helping to fill observational gaps.

The GB survey data will reveal numerous variable stars, including Novae, Cepheids, RR Lyrae, 
and long-period Mira variables. These are standard candles—that is, distance indicators—and 
are therefore useful for studying the structure of the Galaxy \citep{Matsunaga2009, Matsunaga2011, Matsunaga2017}.
In particular, Mira variables are relatively bright and abundant, 
making them excellent tracers of the Galactic structure.
The PRIME GB survey can detect variable stars of intermediate brightness toward the bulge and 
disk regions, complementing existing surveys such as 2MASS, which targets bright variables, 
and VVV, which focuses on faint ones \citep{Sanders2022}.
The Japan Astrometry Satellite Mission for INfrared Exploration (JASMINE) plans to target Mira variables in the Galactic bulge as key scientific objects. JASMINE aims to precisely measure their parallaxes and proper motions to unveil the dynamical structure and evolutionary history of the Galaxy \citep{Kawata2024}. Therefore, identifying suitable Mira variables in advance through the PRIME survey is critically important to maximize the scientific return of the JASMINE mission.

During the off-bulge season, PRIME partner institutions will conduct various scientific observations 
of their interest, such as ToO observations of gravitational wave events, $\gamma$-ray bursts, 
transient searches toward galaxy clusters, and the all-sky grid survey.

%----------------------------------------------------------------------------------------
\acknowledgments
We thank the anonymous referee for the useful suggestions.
The PRIME project is supported by JSPS KAKENHI Grant Number JP16H06287, JP22H00153, JP25H00668,
JP19KK0082, JP20H04754, JP24H01811 and JPJSCCA20210003.
We acknowledge a financial support by Astrobiology Center.
M.T. is supported by JSPS KAKENHI grant No.24H00242.
DPB acknowledges support from NASA grants 80NSSC20K0886 and 80NSSC18K0793.
ET acknowledges financial support from the European Research Council (ERC BHianca, 101002761).

%----------------------------------------------------------------------------------------
\appendix

\section{New Observational strategy}\label{sec-append}

\begin{figure}
\begin{center}
\includegraphics[scale=0.35,keepaspectratio]{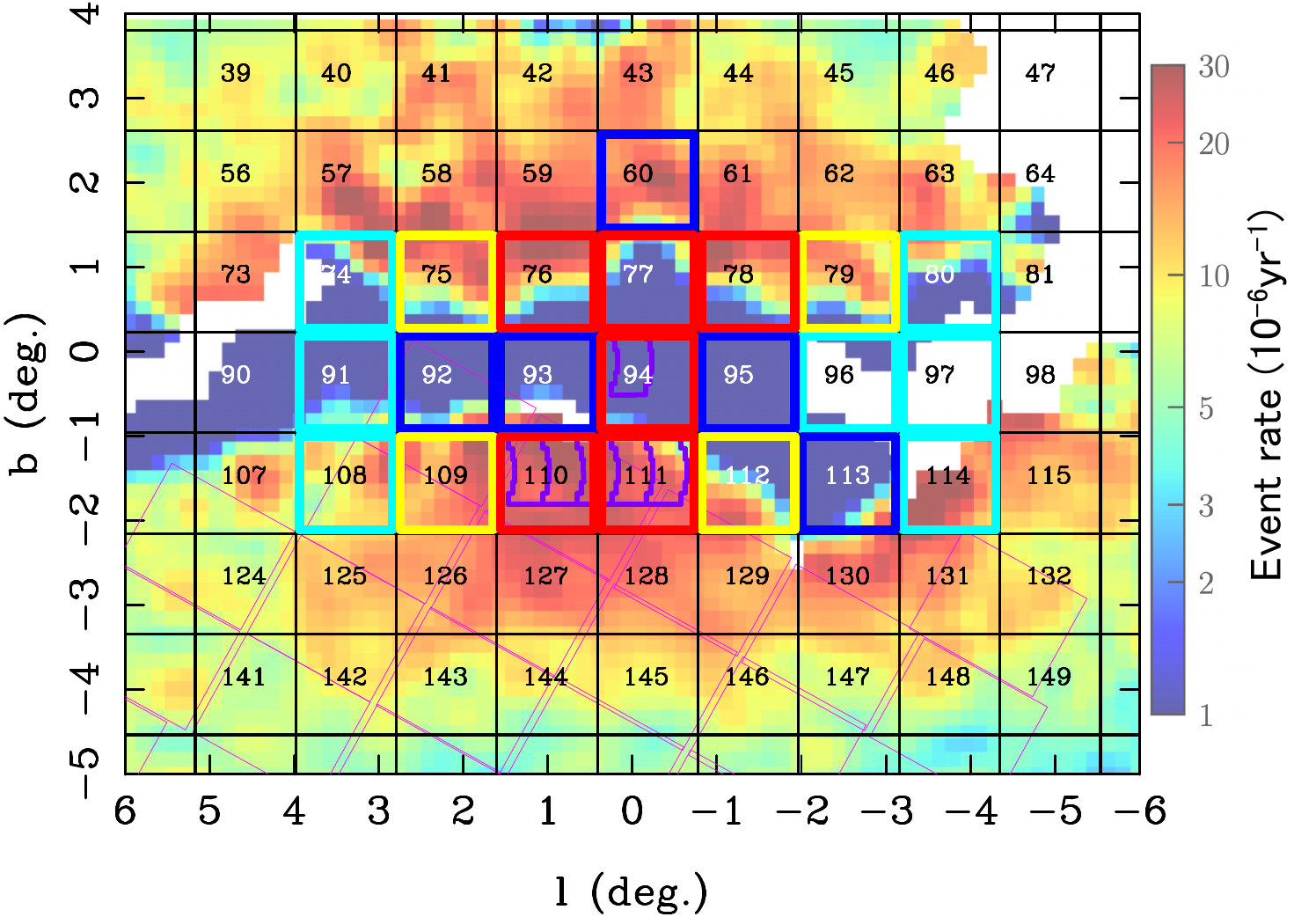}
\caption{
  \label{fig:newfields}
 Same as Figure \ref{fig:fields} for 
new observational fields of PRIME toward the GB since June 2 2025.
The red, yellow, blue and cyan squares represent the fields with cadence of
21.6min, 43.1min, 86.2min and 1day, respectively.
}
\end{center}
\end{figure}

Since June 2 2025, we have modified our observational strategy to increase the cadence by focusing on the Galactic plane, a region that conventional optical surveys cannot access. Figure~\ref{fig:newfields} shows the new PRIME observational fields toward the GB and they are listed in Table \ref{tbl:newfields}. The red, yellow, blue, and cyan squares indicate fields with cadences of 21.6 minutes, 43.1 minutes, 86.2 minutes, and 1 day, respectively. This strategy is similar to that adopted by the MOA survey. The higher cadence allows us to improve the detection efficiency for microlensing exoplanet searches.

%------------------------Table 7.---------------------------------
\begin{deluxetable}{lrr}
\tabletypesize{\scriptsize}
%\rotate
\tablecaption{New observational fields and cadence of PRIME towards the GB since June 2025.
\label{tbl:newfields}
}
\tablewidth{0pt}
\tablehead{
\colhead{ Field } &
\colhead{$l$ ($^\circ$) } &
 \colhead{$b$ ($^\circ$) }
%\colhead{} & 
%\colhead{$(\rm day)$} & 
%\colhead{} & 
%\colhead{} 
}
\startdata
\multicolumn{3}{c}{Cadence: 21.6min (4obs./cycle)} \\
\hline
 GB76 &   1.00681 &   0.82403 \\
 GB77 &  -0.18319 &   0.82403 \\
 GB78 &  -1.37319 &   0.82403 \\
 GB94 &  -0.18319 &  -0.36597 \\
GB110 &   1.00681 &  -1.55597 \\
GB111 &  -0.18319 &  -1.55597 \\
\hline
\multicolumn{3}{c}{Cadence: 43.1min (2obs./cycle)} \\
\hline
 GB75 &   2.19681 &   0.82403 \\
 GB79 &  -2.56319 &   0.82403 \\
GB109 &   2.19681 &  -1.55597 \\
GB112 &  -1.37319 &  -1.55597 \\
\hline
\multicolumn{3}{c}{Cadence: 86.2min (1obs./cycle)} \\
\hline
 GB60 &  -0.18319 &   2.01403 \\
 GB92 &   2.19681 &  -0.36597 \\
 GB93 &   1.00681 &  -0.36597 \\
 GB95 &  -1.37319 &  -0.36597 \\
GB113 &  -2.56319 &  -1.55597 \\
\hline
\multicolumn{3}{c}{Cadence: 1 day} \\
\hline
 GB74 &   3.38681 &   0.82403 \\
 GB80 &  -3.75319 &   0.82403 \\
 GB91 &   3.38681 &  -0.36597 \\
 GB96 &  -2.56319 &  -0.36597 \\
 GB97 &  -3.75319 &  -0.36597 \\
GB108 &   3.38681 &  -1.55597 \\
GB114 &  -3.75319 &  -1.55597 \\
 \enddata
 %\tablenotetext{a}{Likely Brown dwarf lens.}
%\table comments{
%}
\end{deluxetable}
%----------------------------------------------------------------

%----------------------------- FIG. 15 -------------------------------------
%--------------------------------------------------------------------------
%\begin{longrotatetable}
%\begin{deluxetable*}{lllrrrrrrll}
%\tablecaption{Observable Characteristics of 
%Galactic/Magellanic Cloud novae with X-ray observations\label{chartable}}
%\tablewidth{700pt}
%\tabletypesize{\scriptsize}
%\tablehead{
%\colhead{Name} & \colhead{V$_{max}$} & 
%\colhead{Date} & \colhead{t$_2$} & 
%\colhead{FWHM} & \colhead{E(B-V)} & 
%\colhead{N$_H$} & \colhead{Period} & 
%\colhead{D} & \colhead{Dust?} & \colhead{RN?} \\ 
%\colhead{} & \colhead{(mag)} & \colhead{(JD)} & \colhead{(d)} & 
%\colhead{(km s$^{-1}$)} & \colhead{(mag)} & \colhead{(cm$^{-2}$)} &
%\colhead{(d)} & \colhead{(kpc)} & \colhead{} & \colhead{}
%} 
%\startdata
%\enddata
%\end{deluxetable*}
%\end{longrotatetable}

\end{document}